\documentclass[%
 reprint,
superscriptaddress,
 amsmath,amssymb,
 aps,
pra,
]{revtex4-2}

\usepackage{graphicx}
\usepackage{dcolumn}
\usepackage{bm}
\usepackage{dsfont}
\usepackage{physics}
\usepackage[dvipsnames]{xcolor}
\usepackage{mathtools}
\usepackage{cancel}
\usepackage{colortbl}
\usepackage{soul}
\usepackage{siunitx}
\DeclareSIUnit\gal{Gal}
\usepackage{enumerate}
\usepackage{tikz} 
\usetikzlibrary{calc}
\usepackage[normalem]{ulem}
\usepackage[colorlinks,citecolor=blue,linkcolor=magenta,urlcolor=blue]{hyperref} 

\newcommand{\deltatwo}{\delta_\rest}
\newcommand{\deltafour}{\delta_\ie}

\newcommand{\Hfour}{H_\ie}
\newcommand{\Htwo}{H_\rest}
\newcommand{\Hzero}{H_{0}}

\newcommand{\id}{\ensuremath{\mathds{1}}}
\newcommand{\kb}[2]{|#1\rangle\langle#2|}

\newcommand{\diff}{\mathrm{d}}

\newcommand{\cl}{\text{cl}}

\newcommand{\tot}{\text{tot}}

\newcommand{\ent}{\text{ent}}
\newcommand{\ie}{E}
\newcommand{\rest}{m}
\newcommand{\avg}{\text{avg}}
\newcommand{\dipole}{{\text{dipole}}}

\newcommand{\atomdensity}{n}

\usepackage{tikz}
\usepackage{spath3}
\usetikzlibrary{calc,spath3}
\usepackage{pgfplots}
\usepackage{tikz-3dplot}
\usepgfplotslibrary{fillbetween}
\usetikzlibrary{intersections}
\usetikzlibrary{perspective, shadows}

\definecolor{mygray}{RGB}{200, 210, 220}
\pgfdeclareverticalshading{rodshading}{100bp}
{color(0bp)=(black); color(48bp)=(black); color(50.7bp)=(mygray); color(51bp)=(white);  color(51.3bp)=(mygray); color(52bp)=(black);  color(100bp)=(black)}

\pgfdeclareverticalshading{lasershading}{100bp}
{color(0bp)=(black); color(50bp)=(gray!75!black); color(60bp)=(gray!50);  color(70bp)=(gray!75!black); color(100bp)=(black)}

\pgfdeclareverticalshading{lasershadingdark}{100bp}
{color(0bp)=(black); color(50bp)=(gray!50!black); color(60bp)=(gray);  color(70bp)=(gray!50!black); color(100bp)=(black)}

\usepackage{adjustbox}
\pgfplotsset{compat=1.18}

\tikzset{
    partial ellipse/.style args={#1:#2:#3}{
        insert path={+ (#1:#3) arc (#1:#2:#3)}
    }
}

\pgfdeclareverticalshading{mirrorshading}{100bp}
{color(0bp)=(black); color(40bp)=(Goldenrod); color(45bp)=(white); color(50bp)=(Goldenrod); color(60bp)=(Goldenrod); color(61bp)=(white); color(62bp)=(Goldenrod);  color(100bp)=(black)}

\makeatletter
\tikzoption{canvas is plane}[]{\@setOxy#1}
\def\@setOxy O(#1,#2,#3)x(#4,#5,#6)y(#7,#8,#9)%
  {\def\tikz@plane@origin{\pgfpointxyz{#1}{#2}{#3}}%
   \def\tikz@plane@x{\pgfpointxyz{#4}{#5}{#6}}%
   \def\tikz@plane@y{\pgfpointxyz{#7}{#8}{#9}}%
   \tikz@canvas@is@plane
  }
\makeatother

\tikzstyle{reverseclip}=[insert path={(current bounding box.north east) --
  (current bounding box.south east) --
  (current bounding box.south west) --
  (current bounding box.north west) --
  (current bounding box.north east)}
]

\begin{document}

\title{Gravitational Interactions in Nuclear Clocks}
\author{Konstantin Beyer}
\email{kbeyer1@stevens.edu}
\affiliation{Department of Physics, Stevens Institute of Technology, Hoboken, NJ 07030, USA}%
\author{Niklas E. \"{O}nne}
\affiliation{%
  Department of Physics, Stevens Institute of Technology, Hoboken, NJ 07030, USA}
  \affiliation{
  Nordita, KTH Royal Institute of Technology and Stockholm University, SE-114 19 Stockholm, Sweden}
  \affiliation{Department of Physics, Stockholm University, SE-106 91 Stockholm, Sweden}%
\author{Shimon Kolkowitz}
\affiliation{University of California at Berkeley, Department of Physics, Berkeley CA 94720, USA}
\author{Dolev Bluvstein}
\affiliation{Oratomic, Pasadena, CA 91125, USA}
\affiliation{California Institute of Technology, Pasadena, California 91125, USA}
\author{Mikhail Lukin}
\affiliation{Department of Physics, Harvard University, Cambridge, MA 02138, USA}
\author{Jun Ye}
\affiliation{JILA, NIST and University of Colorado, Department of Physics, University of Colorado, Boulder, Colorado 80309, USA }
\author{Igor Pikovski}
\email{pikovski@stevens.edu}
\affiliation{Department of Physics, Stevens Institute of Technology, Hoboken, NJ 07030, USA}
\affiliation{Department of Physics, Stockholm University, SE-106 91 Stockholm, Sweden}

\date{\today}

\begin{abstract}
Atomic clocks achieve extraordinary precision through quantum control over large ensembles of atoms, with precision increasing and atom number advancing steadily. Emerging nuclear clock technology encodes clock states in nuclear rather than electronic transitions, offering a new frontier for precision measurement. Here we study whether a large ensemble of such clock nuclei could be used to observe their mutual gravitational interaction. We investigate the gravitational redshift induced by neighboring atoms, including nuclear energy superpositions as sources of gravity, giving rise to a genuinely quantum gravitational interaction between pairs of nuclei. 
We find that the resulting 
clock frequency shift due to mutual gravity may become non-negligible for a macroscopic number of nuclei, for instance in a solid-state thorium clock, and we propose approaches to mitigate electromagnetic field effects by exploiting spatial symmetry and high-precision spectroscopy.
We further discuss how energy-induced mass superpositions could be directly probed using external gravitational field sensors. By contrast, observing gravity-induced entanglement involving clocks would require substantially greater experimental advances due to decoherence from spontaneous emission.
Taken together, our results suggest that nuclear clocks, either alone or in combination with complementary sensors such as optomechanical systems, may offer a future route toward probing gravity at short distances and testing gravitating quantum source masses beyond the Newtonian limit.
\end{abstract}

\maketitle

\section{Introduction}

The rapid advancement of atomic clocks over recent decades has pushed frequency measurement precision far beyond the requirements of conventional timekeeping, transforming these devices into powerful tools for probing fundamental physics~\cite{ye_essay_2024}.
Modern atomic clocks can measure the redshift of the Earth on a millimeter scale~\cite{bothwellResolvingGravitationalRedshift2022,zhengLabbasedTestGravitational2023}. They are used to constrain violations of Lorentz symmetry and the time variation of fundamental constants~\cite{sanner_optical_2019,godun_frequency_2014,barontini_measuring_2022,porsev_mathrmpr10_2024}, and they have been proposed for the detection of gravitational waves and dark matter~\cite{kolkowitz_gravitational_2016,tino_sage_2019}.
Optical lattice clocks have set new records in precision by combining long atomic coherence times with large numbers of atoms~\cite{oelker_demonstration_2019,bothwellResolvingGravitationalRedshift2022,zhengLabbasedTestGravitational2023}, with the capability of exploring complex many-body phenomena \cite{kolkowitz2017spin,goban2018emergence,aeppli2022hamiltonian}. The lattice clock builds on the idea of trapped ions, where quantum state engineering and preservation of quantum coherence are key~\cite{ludlow_optical_2015}, and complements it with the capability of significantly increasing the number of quantum absorbers~\cite{campbellFermidegenerateThreedimensionalOptical2017}. 

This scalability becomes even more crucial with the advent of nuclear clocks~\cite{peik_nuclear_2003}, which promise dramatically reduced sensitivity to environmental perturbations and potential integration into solid-state platforms~\cite{rellergert_constraining_2010,kazakov_performance_2012}. 
Recent breakthroughs for nuclear clocks include the first resonant laser excitation of the $8.4\,\si{\electronvolt}$ energy gap in the thorium-229 nuclear isomer~\cite{tiedauLaserExcitationTh2292024,elwellLaserExcitationTh2292024} and the demonstration of a prototype nuclear clock~\cite{zhangFrequencyRatio229mTh2024}. Early investigations of systematic effects in solid-state configurations show promising results for scaling the number of clock nuclei while preserving coherence and frequency stability~\cite{higgins_temperature_2025,ooi2026frequency}.
Other potential candidates for a nuclear clock include the metastable excited state of $77\,\si{\electronvolt}$ in uranium-235~\cite{ponceAccurateMeasurementFirst2018} and the $12.4\,\si{\kilo\electronvolt}$ X-ray transition in scandium-45, for which resonant excitation has recently been demonstrated~\cite{shvydkoResonantXrayExcitation2023} and macroscopic numbers of interrogated nuclei---perhaps even reaching Avogadro's number---are envisioned~\cite{peikNuclearClocksTesting2021,roehlsbergerWegZurScandium45Kernuhr2024}.
These platforms are at very different levels of maturity. For thorium-229, the tools required for coherent control of the nuclear states are already in hand. For the more energetic transitions in uranium-235 and scandium-45, no techniques for coherent control currently exist, and a fully scalable nuclear clock in either system is not imminent. Nevertheless, the rapid pace of progress with thorium-229, alongside advances in Mössbauer spectroscopy and free-electron laser technology, provides reason to expect that coherent control of these more challenging transitions may become achievable in the future and scalable nuclear clocks based on them are conceivable.

\begin{figure}
    \centering
    \begin{tikzpicture}
        \node at (0, 0) {\includegraphics[width=\linewidth]{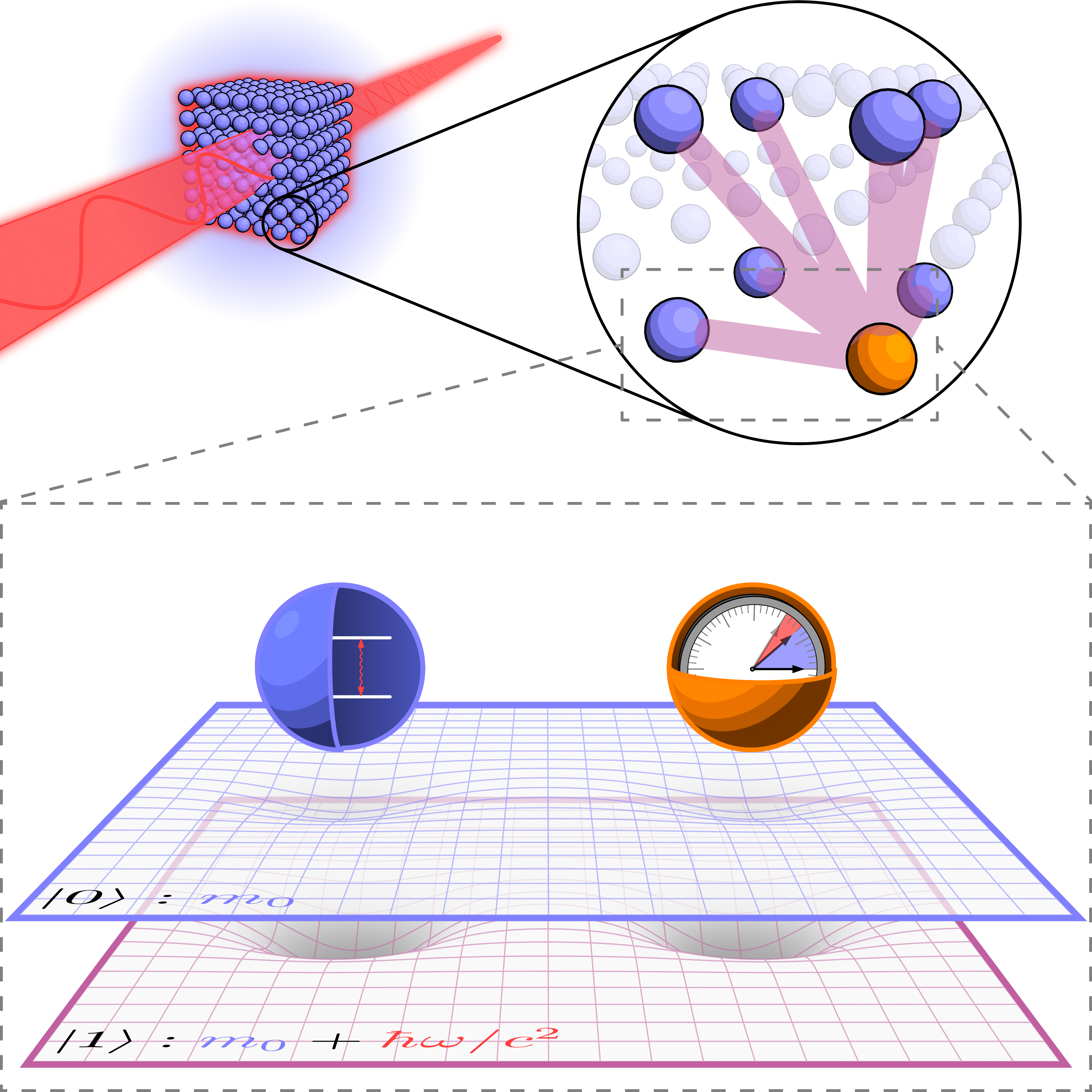}};

        \node (a) at (-3.9, 4) {(a)}; 
        \node (b) at (3.7, 4) {(b)}; 
        \node (c) at (-3.9, -0.1) {(c)}; 


        \node[white, scale=0.75] at (-1.5, -0.55) {$\ket{1}$};
        \node[white, scale=0.75] at (-1.5, -1.35) {$\ket{0}$};
        \node[red!75, scale=0.75] at (-1.25, -0.95) {$\hbar \omega_c$};



        
        \coordinate (clock) at (1.63, -0.97); 

        \node[blue!50, scale=0.9] at ($(clock) + (21:0.9cm)$) {$\delta_m$};
        \node[red!75, scale=0.9] at ($(clock) + (51:0.9cm)$) {$\delta_E$};
        
    \end{tikzpicture}
    \caption{{We investigate the effect of gravitational interactions on the time-keeping of clocks consisting of many atoms with a nuclear clock transition, as shown in (a). (b) Each clock nucleus  experiences the gravitational redshift from all other atoms due to the long-range gravitational interaction. (c) The mutual redshift arises from both the rest mass $m_0$ of the atoms and from the energy superpositions of the clock nuclei, which gravitate according to mass-energy equivalence and add \(m_E = \hbar\omega_c/c^2\) to the total mass if the nucleus is in the \(\ket{1}\) state, creating a stronger gravitational potential. These result in the redshifts $\delta_m$ and $\delta_E$, respectively, from a single neighboring atom. Summing over all individual effects, the total redshift effects  $\Delta_m$ and $\Delta_E$ increase significantly for a macroscopic number \(N\) of clock atoms.}
    }
    \label{fig:clock-grid}
\end{figure}

Nuclear clocks at macroscopic scales open up new possibilities for fundamental physics research. Here, we exploit this potential to study gravity at atomic scales. We investigate the influence of interatomic gravitational effects on clock frequency, encompassing the mutual gravitational redshift from the rest mass of the atoms and from the additional mass associated with the internal energy of the nuclear clock transition~(see Fig.~\ref{fig:clock-grid}). Our results show that the minute gravitational potentials sourced by individual atoms can accumulate to become relevant for macroscopic numbers of clock nuclei at sufficient precision.

We then ask whether the gravity sourced by the clock's energy can be directly measured. Isolating this signal from the overwhelming electromagnetic background is a tremendous experimental challenge: dipole couplings between nuclei are orders of magnitude larger than the expected gravitational interaction between nuclear states, making any measurement within a single clock unrealistic. 
However, seen from the outside, a nuclear clock whose internal states are coherently controlled in superposition behaves as a tunable source mass. We propose two complementary approaches to detect this variable mass: either via the gravitational redshift it imprints on a nearby second clock, or through a force sensor.
Both configurations permit protocols that mitigate the dominant electromagnetic interactions, including by exploiting spatial symmetry.

Measuring such a gravitational effect could serve as the first direct demonstration of a gravitational mass emerging from quantum states of the gravitational charge---gravity sourced by energy in superposition---and would complement the many proposed protocols for future experiments with mechanical systems~\cite{boseBMV2017,marlettoVedralBMV2017,carlesso2019testing,krisnanda2020observable,anastopoulos2020quantum,guff2022optimal,pedernales2022enhancing,kaku2023enhancement,hanif2024testing,toccaceloBenchmarksQuantumCommunication2025a,beyerOnesidedWitnessQuantumness2025}. While our analysis shows that sensing gravitational entanglement with clocks is impractical without substantial progress in suppressing decoherence, our results indicate that detecting superposed gravitational fields created by clocks may become feasible with future advances in nuclear clock technology.

The remainder of the paper is organized as follows. We first set up the case of two clock atoms in Section~\ref{sec:two-atoms} and calculate the emerging gravitational effects arising from both the rest mass and the energy of the clock transition.
We then turn to the case of \(N\) atoms and calculate the scaling of the gravitational influence with the number \(N\).
We provide concrete numbers for the required averaging time to resolve such effects in solid-state nuclear clocks. 
In Section~\ref{sec:experimental}, we 
focus on the question of how to probe the gravitational field sourced by the superposed internal degrees of freedom with a second clock or a force sensor.
We then compare gravity sourced by energy superpositions with gravity sourced in other physical implementations in Section~\ref{sec:compare}, and highlight the unique aspects offered by nuclear clocks. 

\section{Gravitational interactions in atomic clocks}
\label{sec:two-atoms}
We model atomic clocks (and equally nuclear clocks) as two-level systems with rest mass $m_0$ and an addressable energy transition of $E=\hbar \omega_c$.
Each single clock atom is described by a Hamiltonian of the form
\begin{align}
\label{eq:bare-Hamiltonian}
    H = H_r + H_c = m_0 c^2 \id + \hbar\omega_c \kb{1}{1} .
\end{align}
The subscript \(r\) denotes the rest mass part of the Hamiltonian, which is a scalar, and \(c\) denotes the actual \emph{clock} part of the Hamiltonian. {The ground state energy is included in the rest mass $m_0$.} As long as gravitational interactions are neglected, only the bare clock Hamiltonian \(H_c\) leads to observable effects, since the rest mass part \(H_r\) is proportional to the identity and therefore only contributes a global phase. 

Let us now consider two of these clock atoms at a distance \(a\) from each other. For the sake of clarity, we assume throughout that all atoms are completely fixed to their respective positions. Thus, we omit any effects that result from the motional degrees of freedom. {Our analysis builds on the study of proper time evolution of quantum composite systems \cite{Zych2011,sinha2011atom, Pikovski2015,zych2016general,orlando2017does,zhou2018quantum,loriani2019interference,roura2020gravitational,paige2020classical, khandelwal2020universal,hartong2024coupling,barzel2024entanglement,borregaard2024testing, covey2025probing}, where clocks in superposition of different trajectories experience time dilation from a background field. The changes in proper time can be equivalently described in terms of the mass-energy equivalence of clock atoms, where the internal states result in different masses \cite{Zych2011,pikovski2017time,sonnleitner2017will,anastopoulos2018equivalence,zych2019gravitational,schwartz2019post,martinez2022ab}. This perspective allows one to easily extend the scenario to the mutual gravitational interactions of clock atoms~\cite{castroruizEntanglementQuantumClocks2017}, described by the Hamiltonian}
\begin{align}
\label{eq:gravitational-Hamiltonian}
    H_{\text{int}}= -\frac{G}{c^4 a} H^A \otimes H^B,
\end{align}
where \(G\) is the gravitational constant.
This Hamiltonian emerges from the fact that the mass of the atom is equivalent to its total energy. An atom in the ground state has the rest mass \(m_0\), while the same atom in the excited state (which can be a nuclear excitation) has a mass \(m_1 = m_0 +  \hbar\omega_c / c^2\) and therefore sources a stronger gravitational potential than the ground state.
The Newtonian gravitational interaction energy between two arbitrary masses \(m_A\) and \(m_B\) is given by $\Phi = -(G m_A m_B)/{a}$.
Thus, neglecting the external degrees of freedom of the atoms and fixing their distance to \(a\), the Newtonian gravitational interaction energy only depends on the internal states of the atoms, which directly leads to the interaction Hamiltonian in Eq.~\eqref{eq:gravitational-Hamiltonian}. 

The total Hamiltonian for the joint system of two atoms is then given by the local bare Hamiltonians \(H\) in Eq.~\eqref{eq:bare-Hamiltonian} and their gravitational interaction \(H_\text{int}\) in Eq.~\eqref{eq:gravitational-Hamiltonian}. 
Expanding in orders of \(c^{-2}\) and omitting terms proportional to the physically irrelevant global identity we find \(
    H_\tot = \Hzero + \Htwo + \Hfour\),
with \(\Hzero= H_c^A + H_c^B,\) and
\begin{align}
           \Htwo =- \frac{Gm_0}{c^2 a} \left(H_c^A + H_c^B\right),&&
  \label{eq:Hquantum}       &\Hfour = -\frac{G}{c^4 a} H_c^A\otimes H_c^B.
\end{align}
The term \(\Htwo\) to order \(c^{-2}\) contributes a {local} redshift of each individual atom's clock frequency due to the classical background gravitational field sourced by the other atom's \emph{rest mass} \(m_0\). This part is essentially classical.
The actual quantum Hamiltonian \(\Hfour\) due to the interaction of the \emph{internal energies} is of order \(c^{-4}\){, and is a direct consequence of the fact that an energy superposition sources a gravitational field in superposition.}

\subsection{Clock evolution due to the mutual gravitational interaction}\label{sec:time-evolution}
In a standard atomic clock experiment, all atoms are initially prepared in a product state and the readout measurements are local observables. 
Thus, to analyze the observable gravitational effects for these standard scenarios, it suffices to calculate the reduced dynamics of the internal degrees of freedom of individual atoms, i.e., we can regard each single clock atom as an open quantum system whose environment is constituted by the other atom.
The {exact} dynamics of atom \(A\) is given by
\begin{align}
\label{eq:reduced-dynamics}
    \rho_A(t) = \tr_B\left[e^{-\frac{i}{\hbar} H_\text{tot} t}  (\rho_A\otimes\rho_B) e^{\frac{i}{\hbar} H_\text{tot} t}\right],
\end{align}
where \(\rho_{A,B}\) are the initial states of the respective subsystems.
The impact of the gravitational interaction can most easily be understood by considering the resulting non-Markovian master equation for the dynamics of atom \(A\) (see Appendix~\ref{app:master-equation} for the derivation): 
\begin{align}
\label{eq:master-equation}
    \dot\rho_A = -\frac{i}{\hbar} (1 - \delta_t) [H_c,\rho_A] + \Gamma_t (\sigma_z \rho_A \sigma_z - \rho_A),
\end{align}
where \(\delta_t\) and \(\Gamma_t\) are time-dependent functions that describe a frequency shift and a dephasing rate, respectively. 

The Hamiltonian part of Eq.~\eqref{eq:master-equation} shows that the frequency of the clock changes as
\begin{align}
\label{eq:single-atom-omega-shift}
    \omega_c \mapsto (1 - \delta_t)\omega_c. 
\end{align}
{This shift captures that gravity effectively alters the ticking rate of the bare clock due to the mutual redshift.} 
Up to order \(c^{-4}\), the shift is time-independent, \(\delta_t = \delta\), and given by
\begin{align}
\label{eq:single-atom-frequency-shift}
   \delta &= \deltatwo + \deltafour + O(c^{-12}),
\end{align}
where  
\begin{align}
\label{eq:delta-rm}
\deltatwo &=  \frac{G m_0}{c^2 a}
\end{align}
describes the classical redshift of the frequency of atom \(A\) due to the rest mass \(m_0\) of atom \(B\) at distance \(a\), and
\begin{align}
\label{eq:delta-qu}
\deltafour &=  \frac{G}{c^4 a} \tr[H_c\rho_B] =  \frac{G}{c^2 a} \frac{\langle E_B \rangle}{c^2}
\end{align}
is the frequency shift caused by the quantum interaction Hamiltonian \(\Hfour\), which can be interpreted as a redshift due to the mass that is equivalent to the internal {average} energy \(\langle E_B \rangle\) of atom \(B\).

Unlike the constant contribution \(\deltatwo\), the shift \(\delta_\ie\) can in principle directly be measured by varying the initial state of the clock. 
Let us consider an initial product state of the two atoms of the form 
\begin{align}
\label{eq:initial-state-alpha}
    \ket{\psi} = \left[ \cos \left(\frac{\alpha}{2}\right) \ket{0} + \sin \left(\frac{\alpha}{2}\right)\ket{1}  \right]^{\otimes 2}.
\end{align}
Since the frequency shift \(\deltafour\) of atom \(A\) depends on the internal energy of atom \(B\) and vice versa, we have 
\begin{align}
\label{eq:delta4-alpha}
    \deltafour(\alpha) = \frac{G \hbar \omega_c}{c^4 a} \frac{1-\cos \alpha}{2}.
\end{align}
{The energy-induced redshift is thus dependent on the initial preparation through $\alpha$. In contrast,} the bare frequency of the clock \(\omega_c\) and the rest-mass-induced redshift \(\deltatwo\) are independent of \(\alpha\). Therefore, running the clock experiment with different \(\alpha\) could in principle isolate the redshift due to the neighboring clock atoms' internal energy
and thus post-Newtonian gravity sourced by microscopic systems.

The non-unitary dephasing dynamics given by the second part of Eq.~\eqref{eq:master-equation} results from a buildup of correlations between atom \(A\) and \(B\) due to the gravitational interaction Hamiltonian \(H_\ie\) in Eq.~\eqref{eq:Hquantum}. This effect was first proposed in Ref.~\cite{castroruizEntanglementQuantumClocks2017}.
Unlike the frequency shift, for the dephasing rate the first relevant order is \(c^{-8}\):
\begin{align}
\label{eq:depahsing-first-relevant-order}
    \Gamma_t = \frac{G^2 \omega_c^2 t}{2 a^2 c^8}\left(\tr[\rho_B H_c^2] - \tr[\rho_B H_c]^2\right)+O(c^{-16}).
\end{align}
We see that the strength of the dephasing of \(A\) depends on the energy variance in \(B\). If both clock atoms start in a pure energy superposition state \(\ket{\psi}\) such as in Eq.~\eqref{eq:initial-state-alpha}, the dephasing can be attributed to a buildup of entanglement between the systems.
In Appendix~\ref{app:dephasing}, we generalize the results of Ref.~\cite{castroruizEntanglementQuantumClocks2017} and reevaluate the numerical estimates by accounting for a realistic spatial distribution of {\(N\)} atoms.
While the decoherence rate grows with the number of clock atoms, 
we find that even for macroscopic \(N\) the resulting effect is negligibly small, as the 
decoherence time exceeds the age of the universe by far. 
We therefore ignore it in what follows and focus only on the frequency shift \(\delta_t\) in the Hamiltonian part of Eq.~\eqref{eq:master-equation}, which can become observable. 

\subsection{\(N\)-atom frequency shift}
While the mutual gravitational shift for two atoms is entirely negligible, including more clock atoms can significantly increase the effect, {due to the universally attractive nature of gravity}. 
The total Hamiltonian for the joint system of \(N\) atoms reads
\begin{align}
\label{eq:total-Hamiltonian}
    H_\text{tot} =& \Hzero + \Htwo + \Hfour  \notag\\
    =& \sum_i H_c^i - \frac{G m_0}{c^2} \sum_i \sum_{j\neq i} \frac{1}{r_{ij}}H_c^i \notag\\
    &-\frac{G}{c^4} \sum_i \sum_{j<i} \frac{1}{r_{ij}} H_c^i\otimes H_c^j,
\end{align}
where the first term \(\Hzero\) represents the local clock Hamiltonians without gravitational interaction. The second one \(\Htwo\) of order \(c^{-2}\) is the classical redshift of atom \(i\) due to the presence of the rest masses of the other atoms, and the last term \(\Hfour\) is the mutual {energy-induced} {gravitational} interaction between each pair of atoms which is of order \(c^{-4}\).
The atomic distances are denoted by \(r_{ij}\). 

{The resulting dynamics is again described by the master equation \eqref{eq:master-equation}, except that a single clock atom \(i\) now experiences contributions of the other \(N-1\) atoms, over which the trace is performed. } Let us consider an initial product state of the form
\begin{align}
\label{eq:initial-product-state}
    \ket{\psi} = \left[ \cos \left(\frac{\alpha}{2}\right) \ket{0} + \sin \left(\frac{\alpha}{2}\right)\ket{1}  \right]^{\otimes N}.
\end{align} 
The redshifts due to the \(N-1\) surrounding atoms add up, and the frequency of atom \(i\) is given by \(
    \omega_i = \omega_c(1 - \Delta_i)\),
with \(\Delta_i = [\delta_\rest + \delta_\ie(\alpha) ]\Xi_i\),
where \(\delta_{\rest,\ie}\) are the shifts for two atoms at distance \(a\) as given in Eqns.~(\ref{eq:delta-rm}, \ref{eq:delta4-alpha}) and
\begin{align}
\label{eq:Xi-i}
    \Xi_i = \sum_{j\neq i} \frac{a}{r_{ij}} 
\end{align}
is the sum over the inverse distances between atom \(i\) and all other atoms, normalized by the nearest neighbor distance \(a\).
The sum depends on the geometry of the atomic ensemble and the total atom number \(N\). 
Due to the different positions of the atoms in the ensemble, the shift \(\Delta_i\) will differ slightly from atom to atom. An atom at the center experiences a larger redshift than an atom at the boundary. 

We calculate the geometric factor \(\Xi_i\) for a spherical shape of an atomic ensemble,  with approximately constant atom number density \(\atomdensity = N/V \approx a^{-3}\).
For \(N\gg1\), we can approximate the sum by an integral over the atom density \(\atomdensity\).
The  radius of the sphere is 
\begin{align}
\label{eq:radius}
    R = \left(\frac{3}{4\pi\atomdensity} N \right)^{1/3} = a\left(\frac{3}{4\pi} N \right)^{1/3}.
\end{align}
The geometric factor for an atom \(i\) at distance \(l\) from the center of the sphere then reads
\begin{align}
    \Xi(l) &= a \int\limits_0^R \!\!\diff r \!\!\int\limits_0^\pi \!\! \diff \theta \!\! \int\limits_0^{2\pi} \!\! \diff \phi \,\frac{\atomdensity r^2 \sin \theta}{\sqrt{r^2 + l^2 - 2rl \cos\theta}} \notag\\
    & = \frac{2 \pi a}{3}\left(3 R^2 - l^2  \right) \atomdensity. 
\end{align}
For an atom at the boundary of the sphere \(l=R\), whose redshift should be the smallest, we have
\begin{align}
    \Xi_\text{min} = \frac{4 \pi a}{3} R^2 \atomdensity = \left(\frac{4\pi}{3} \right)^{1/3} N^{2/3}, 
\end{align}
whereas for an atom at the center of the sphere \(l=0\) we find
\begin{align}\label{eq:Xi-max}
    \Xi_\text{max} &= 2\pi a R^2 \atomdensity = \frac{3}{2}\left(\frac{4\pi}{3} \right)^{1/3} N^{2/3}  = \frac{3}{2} \Xi_\text{min}.
\end{align}
Averaging over all atoms in the volume \(V\) of the sphere we get
\begin{align}
\label{eq:Xi-avg}
    \Xi_\avg &= \frac{1}{V} \int_0^R 4\pi l^2 \Xi(l) \,\diff l \notag \\
    &=\frac{8\pi a}{5} R^2 \atomdensity = \frac{6}{5 } \left(\frac{4\pi}{3} \right)^{1/3} N^{2/3}.
\end{align}
Thus, the shift measured by a clock that averages over the read-out of all atoms is given by
\begin{align}
\label{eq:Delta-avg}
    \Delta_\avg = \Delta_\rest + \Delta_\ie = [\deltatwo + \deltafour(\alpha)]\Xi_\avg.
\end{align}
{Importantly, the resulting frequency shift is now enhanced by $N^{2/3}$, which can be substantial for a macroscopic number of clock {atoms} as envisioned for nuclear clocks. Note that for simplicity, we have averaged over all atoms in the sample. {Since} each {atom} experiences a slightly different redshift, {depending on its specific position, this} averaging would  {cause a small} dephasing effect. {However,} if individual atoms can be spatially resolved, as has been demonstrated in Ref.~\cite{marti2018imaging} for optical lattice clocks, this effect could in principle be compensated.}

\subsection{Scaling of the gravitational signature}

Increasing the number of atoms in the clock not only increases the strength of the gravitational effect as shown above, but it also reduces the quantum projection noise of the read-out and therefore improves the precision of the clock.
This precision is fundamentally limited by
\begin{align}
\label{eq:sigma}
    \sigma = \frac{1}{2\pi f \sqrt{N} \sqrt{\tau T}},
\end{align}
where \(f = \omega_c/(2\pi)\) is the frequency of the clock transition, \(\tau\) is the lifetime of the transition and \(T\) is the total averaging time. 
In order for the collective rest-mass-induced shift $\Delta_{\rest}$ and the energy-induced shift $\Delta_{\ie}$ from Eq.~\eqref{eq:Delta-avg} to become relevant for the clock, they have to be at the level of the clock precision, i.e.\ 
\begin{align}
\label{eq:sigma-delta}
    \sigma = \Delta_{\rest,\ie}.
\end{align}
Since the number of atoms \(N\) contributes both to the sourcing and the sensing of the signal, its scaling is more favorable than the scaling in the averaging time \(T\).
Solving Eq.~\eqref{eq:sigma-delta} for \(T\) as a function of \(N\), we find
\begin{align}
\label{eq:required-T}
    T_{\rest,\ie} = \left(\frac{5}{12\pi} \sqrt[3]{\frac{3}{4\pi}}\right)^{2} \frac{1}{\delta^2_{\rest,\ie} f^2 \tau N^{7/3}},
\end{align}
with the shifts \(\delta_{m,E}\) for a single pair of atoms as given in Eq.~\eqref{eq:delta-rm} and Eq.~\eqref{eq:delta4-alpha}, respectively.

\subsection{Strength of mutual gravity in clocks}
\label{sec:strength}
It has been proposed that solid-state nuclear clocks might reach macroscopic atom numbers in the future~\cite{roehlsbergerWegZurScandium45Kernuhr2024}.
The strong scaling for \(N\) in Eq.~\eqref{eq:required-T} makes the mutual gravitational redshift a conceivable observable in future experiments, or will at least impose some limitations on the clock precision for sufficiently large atom numbers, which we now estimate. 

To date, the most promising candidates for such clocks are based on nuclear transitions in thorium-229, and possibly in uranium-235 and scandium-45 in the future. 
While thorium has an $8.4$\,eV clock transition in the optical regime corresponding to a wavelength $\lambda_c = 148$\,nm, scandium has a $12.4$\,keV X-ray energy gap. Thus, for the frequency shift \(\Delta_\ie\), which depends on the internal energy, the scandium case is more promising. 
The metastable uranium-235 isomer lies in between with 77\,eV, but has an extremely long half-life of 26~min~\cite{browneNuclearDataSheets2014}.
In the following, we will estimate concrete numbers for the strength of the gravitational shifts for these examples.
The relevant parameters are the rest mass of a single atom \(m_0\), the wavelength of the clock transition \(\lambda_c = 2 \pi c /\omega_c = c/f\), and the atom density \(\atomdensity\). 
We assume a solid-state configuration and set the distance \(a\) to be the edge length of a simple cubic unit cell, i.e., $a = \atomdensity^{-1/3}$.
In experiments, the clock atoms will typically be embedded into a crystal such as \(\text{CaF}_2\) or come in a crystallized molecular form, e.g.\ \(\text{ScF}_3\)~\cite{elwellLaserExcitationTh2292024, kuznetsovaSpectralFluxEnhancement2024}.
This will change the effective rest mass \(m_0\) and the relevant distance \(a\) but does not affect the overall scaling, so we will neglect it here for simplicity.
In Table~\ref{tab:summary}, we show the relevant parameters of all three materials.

\def\arraystretch{1.5}
\begin{table}
    \centering
    \begin{ruledtabular}
    \begin{tabular}{c|c|c|c}
    & Thorium-229 & Scandium-45 & Uranium-235 \\
    \hline \hline
    \(m_0\)  [kg] & \(3.8 \times 10^{-25}\,\) & \(0.75 \times 10^{-25}\) & \(3.9 \times 10^{-25}\) \\
    \(\lambda_c\)  [nm] & 148 & 0.1 & 16\\
    \(\atomdensity\) [cm\(^{-3}\)] & \(3.2 \times 10^{22}\)  & \(3.1\times 10^{22}\) & \(5.0\times 10^{22}\) \\
    \(a\) [pm]  & 320 & 315 & 271\\
    \(\tau\) [s] & 641 & 0.47 & 2251\\
    \hline \hline 
    %
    \(T_\rest(N=N_A)\) & 100\,ns & 1.7\,ns & 0.4\,ns\\
    \(N_\rest(T=100\)\,h) & \(10^{18}\) & \(10^{18}\) & \(10^{17}\)\\

    \hline\hline
        \(T_\ie(N=N_A)\) & 9\,Myr & 21\,h & 357\,yr\\
        \(N_\ie(T=100\)\,h) & \(10^{28}\) & \(10^{23}\) & \(10^{26}\)
    \end{tabular}
    \end{ruledtabular}
    \caption{
    The parameters for the three possible solid-state nuclear clocks: \(m_0\) is the rest mass of a single atom, \(\lambda_c\) the wavelength of the clock transition, \(\tau\) its lifetime, \(n\) the atom density, and \(a\) the nearest neighbor distance between two atoms.    
    The numbers \(T_{\rest,\ie}\) and \(N_{\rest,\ie}\) show under which conditions the gravitational redshifts reach the fundamental limit of the clock's precision (see Eq.~\eqref{eq:sigma-delta}).
    The relationship between averaging time and atom number is given in Eq.~\eqref{eq:required-T}.
    For an averaging time \(T=100\,\)h, the rest mass shift becomes relevant for atom numbers \(N_\rest\) well below Avogadro's.
    By contrast, the internal energy shift would only be resolvable in scandium as can be seen from the required averaging time \(T_\ie\), given a macroscopic atom number. 
    }
    \label{tab:summary}
\end{table}

\subsubsection{Rest mass redshift}

Setting \(N = N_\text{A}=6.022\times 10^{23}\) as a benchmark for a macroscopic ensemble, we calculate the averaging time \(T_m\) required for the rest mass effect to become relevant [see Eq.~\eqref{eq:required-T}]. 
We find values in the nanosecond regime (see Table~\ref{tab:summary}). 
The rest mass of the clock thus has a non-negligible influence on its own frequency in macroscopic nuclear clocks.
To estimate the minimum number of atoms needed for this effect to have an influence, we fix the averaging time to \(T = 100\)\,h---roughly the order of magnitude achieved today---and solve Eq.~\eqref{eq:required-T} for \(N\).
The rest mass shift becomes relevant at atom numbers as low as \(N_m \sim 10^{18}\) in thorium, with similar numbers for scandium and uranium (see Table~\ref{tab:summary}), demonstrating that this gravitational influence plays a role well below Avogadro's number and will have to be taken into account as a systematic shift in such nuclear clocks, including the thorium-based nuclear clock now in rapid development.

\subsubsection{Internal energy redshift}
\label{sec:internal-shift}

Resolving the much weaker, but fundamentally relativistic, shift due to the gravitational effect of the clock's internal energy is considerably harder. 
For \(N = N_A\) atoms, the required averaging time for scandium is roughly \(21\)\,h, whereas thorium would require millions of years and uranium at least centuries (see Table~\ref{tab:summary}). This stark discrepancy is readily understood from Eq.~\eqref{eq:required-T}: the pairwise shift \(\delta_\ie\) depends on the transition frequency through mass-energy equivalence [see also Eq.~\eqref{eq:delta4-alpha}], so the required averaging time scales as \(T_\ie \sim f^{-4}\), explaining why the high-energy scandium transition is so much more favorable. 
For completeness, we also list in Table~\ref{tab:summary} the required atom numbers for \(T=100\)\,h, showing that thorium and uranium would require numbers well beyond Avogadro's which seems practically impossible.

\section{Prospects for experimental realization}
\label{sec:experimental}

The results above show that as nuclear clock technology matures, the interatomic gravitational interactions may become non-negligible for macroscopic atom numbers in some systems. Nevertheless, extracting experimentally the purely gravitational signal would be a much bigger challenge, given the plethora of other, stronger effects in the system. We now discuss how one may in principle approach this challenge, building on high-precision spectroscopy techniques and the control over the system.
Specifically, the energy-dependent total mass of the clock opens the possibility to selectively change the gravitational signal. Together with the abilities to modify the geometry, separate the sources from the probes, and apply quantum control on both, it may become possible to reject systematic errors from the deleterious electromagnetic background while isolating the gravitational signal, similarly to how searches for physics beyond the Standard Model are conducted in atomic and molecular systems \cite{ye_essay_2024,RevModPhys.90.025008}.

Let us for simplicity assume a spherical clock ensemble again. 
The total gravitational potential  produced by the clock atoms and their internal energy is given by 
\begin{align}
    \Phi(r,\alpha) = -\frac{G}{r}\left[M_0 + M_E(\alpha) \right],
\end{align}
where \(r\) is the distance from the center of the sphere, \(M_0\) is the rest mass of the clock consisting of $N_S$ atoms, and 
\begin{align}\label{eq:total-energy-mass}
    M_\ie(\alpha) = \frac{\hbar \omega_c N_S}{c^2} \frac{1-\cos{\alpha}}{2}
\end{align}
is the superposition-dependent effective mass of the internal energy, as the parameter $\alpha$ captures the initial clock state through Eq.~\eqref{eq:initial-product-state}. In the following we will neglect the fully classical and static rest mass offset \(M_0\) and only focus on the question of how to detect the gravitational field sourced by \(M_E\). 
Thus the relevant mass in Eq.~\eqref{eq:total-energy-mass} is the \textit{active} mass associated with the clock's energy levels \cite{will1976active,treder2013fundamental,fragkos2025probing}.
For \(0 < \alpha < \pi \), the internal states are in an energy superposition. 
Thus, measuring the \(\alpha\)-dependent gravitational field created by the nuclear states would be a direct demonstration of gravity sourced by superposed energies.

\bigskip

\subsection{Measuring superpositions of the gravitational field with clocks}
\label{sec:shell}
\begin{figure}[t!]
	\centering
    \begin{tikzpicture}
        \node at (0, 0) {\includegraphics[width=0.75\columnwidth]{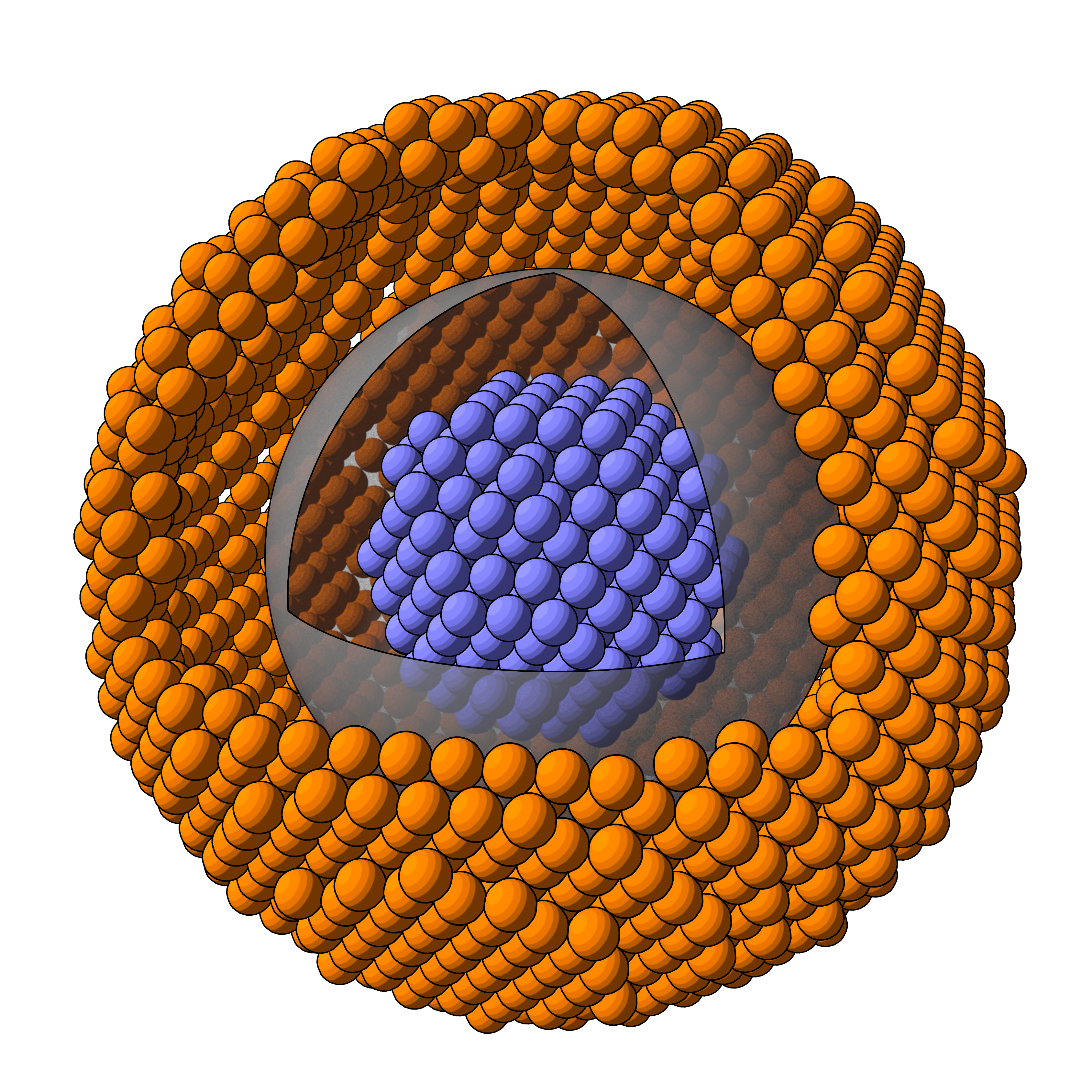}}; 
        


    \end{tikzpicture}
	\caption{Two nuclear clock systems, each consisting of a macroscopic number of atoms, will induce a mutual redshift on similar scales as the single-clock-system setup. Here, one clock (blue) can serve as the source, while the other (orange) as the probe of the gravitational field. This allows for direct modulation of the source clock's gravitational field induced by its internal levels, and read-out with lock-in detection which can mitigate the inevitable background noise within one clock sample. 
    The spherical geometry maximizes the interaction strength and can be utilized to mitigate dipole effects. Additionally, a Faraday shield between the two clocks could be used to further isolate the gravitational shift from electromagnetic forces.
    }
	\label{fig:shell-setup}
\end{figure}

We first consider how to exploit the geometry of the source and probe. One can separate clocks used to source the gravitational field from those used solely to probe the resulting field. Using an external probe avoids the inherent electromagnetic noise floor  within a single clock setup, such as from van der Waals forces, magnetic quadrupole moments, or even the long-range cooperative Lamb shift \cite{hutsonObservationMillihertzlevelCooperative2024}. 
The fact that the gravitational signal is due to a monopole superposition (i.e.\ the charge itself is in superposition rather than the location) means that specific geometric arrangements can help mitigate other effects, which are dipolar or higher multipoles. Maintaining a spherical geometry is then particularly useful.

Let us consider a scenario with two sets of clock atoms as depicted in  Fig.~\ref{fig:shell-setup}.
The inner clock sphere of radius \(R\) serves as the source of gravity, and can experimentally be tuned by changing the superposition parameter \(\alpha\) of its nuclear states, as given in Eq.~\eqref{eq:total-energy-mass}.
The outer shell, which is described by an average radius \(L\) and a thickness \(d\), consists of clock atoms that measure the resulting redshift from the inner clock. The two ensembles need not be the same, and using different species might offer advantages such as for suppressing the radiative dipole-dipole coupling between the source and the probe atoms, and the ability to address the systems separately.

The concentric and separated geometry is advantageous for the suppression of electromagnetic influences which easily mask any gravitational effect.
Before we look in more detail at suitable protocols that help mitigate this noise, we estimate the strength of the gravitational signature in this setup.

Due to Gauss' theorem, the total redshift sourced by the inner clock sphere is the same as if the mass was concentrated at the center. 
All sensor clock atoms in the shell see a slightly different redshift because of their different distances to the source. However, since the thickness of the shell will usually be much smaller than the radius, this inhomogeneous broadening is much smaller than the one in the single spherical clock ensemble discussed before [see Eq.~\eqref{eq:Xi-max}] (or could be avoided altogether by individually addressing the clock atoms or entangling atoms as described in Appendix~\ref{app:coherent-cancelling}).
The average redshift seen by the clock atoms in the shell is given by
\begin{align}
\label{eq:Delta-avg-shell}
    \Delta_\avg(\alpha) &=  \frac{G M}{c^2 V} \int_{V} \frac{1}{r} \diff V = \frac{12 G L M_\ie(\alpha)}{c^2(12 L^2 + d^2)},
\end{align}
where $M_\ie(\alpha)$ is again given by Eq. \eqref{eq:total-energy-mass}, \(V\) is the volume of the shell and \(r\) is the distance of an individual atom in the shell from the center of the sourcing sphere.
Note that the parameter \(\alpha\) controls the mass of the sourcing sphere and the amplitude of the mass change is given by \(M_E(\alpha=\pi/2)\).
The initial state of the probing shell is not varied, as this would reintroduce varying electromagnetic contributions in the probing clock. The internal energy of the shell only contributes an irrelevant constant offset, while the signature of interest is the varying redshift due to the variation of \(\alpha\) in the source. 
The scaling in the number of atoms \(N\) is essentially the same as in the case of a single clock. Assuming \(N_S = N_P= N=6\times 10^{23}\) scandium atoms in both the source sphere and the probe shell, we arrive at a total redshift of 
\(\Delta_\avg \approx 10^{-34}\) and a necessary averaging time of \(T_E = 1.7\) days.
The only numerical difference to the case discussed in Section~\ref{sec:internal-shift} is the mean distance of the sensing clock atoms to the source of gravity, which is slightly greater here and explains the increased averaging time in comparison to the value in Table ~\ref{tab:summary}.

Techniques used in high-precision spectroscopy such as double electron-electron resonance (DEER) can be used for optimized measurements in coupled controllable systems, combined with control-sequences such as dynamical decoupling (DD). Here we discuss how one might apply these techniques for rejecting systematic noise during the measurement of the mutual gravitational interactions, assuming the same tools will become available for nuclear clocks as currently employed for sensing with color centers and atomic clocks.

\begin{figure}[t!]
	\centering
    \begin{tikzpicture}[xscale=0.8]

        \begin{scope}[yshift=1.1cm] 

            \pgfmathsetmacro{\lw}{\linewidth/1cm}
            \pgfmathsetmacro{\fwhm}{0.22}
            \begin{scope}
                \clip (-0.5\linewidth, -1) rectangle (0.5\linewidth, 1);
                \draw[red!75, fill=red!75, fill opacity=0.5, line width=1pt, domain=-0.5*\lw:0.5*\lw, samples=300] plot[smooth] (\x, {0.4*exp(-(\x+0.35*\lw)*(\x+0.35*\lw)/((0.5*\fwhm)^2)) + 0.4*exp(-\x*\x/(\fwhm^2)) + 0.4*exp(-(\x-0.35*\lw)*(\x-0.35*\lw)/((0.5*\fwhm)^2))-1});
            \end{scope}
        
            \draw[gray, line width=0.75pt] (-0.5\linewidth, -1) rectangle (0.5\linewidth, 1); 

            \node[right=1mm] at (-0.35\linewidth, -0.65) {$\pi/2$}; 
            \node[right=2mm] at (0, -0.65) {$\pi$}; 
            \node[right=1mm] at (0.35\linewidth, -0.65) {$\pi/2$}; 

            \node[inner sep=0pt] (probe) at (-0.63\linewidth, 0.2) {\includegraphics[width=1cm]{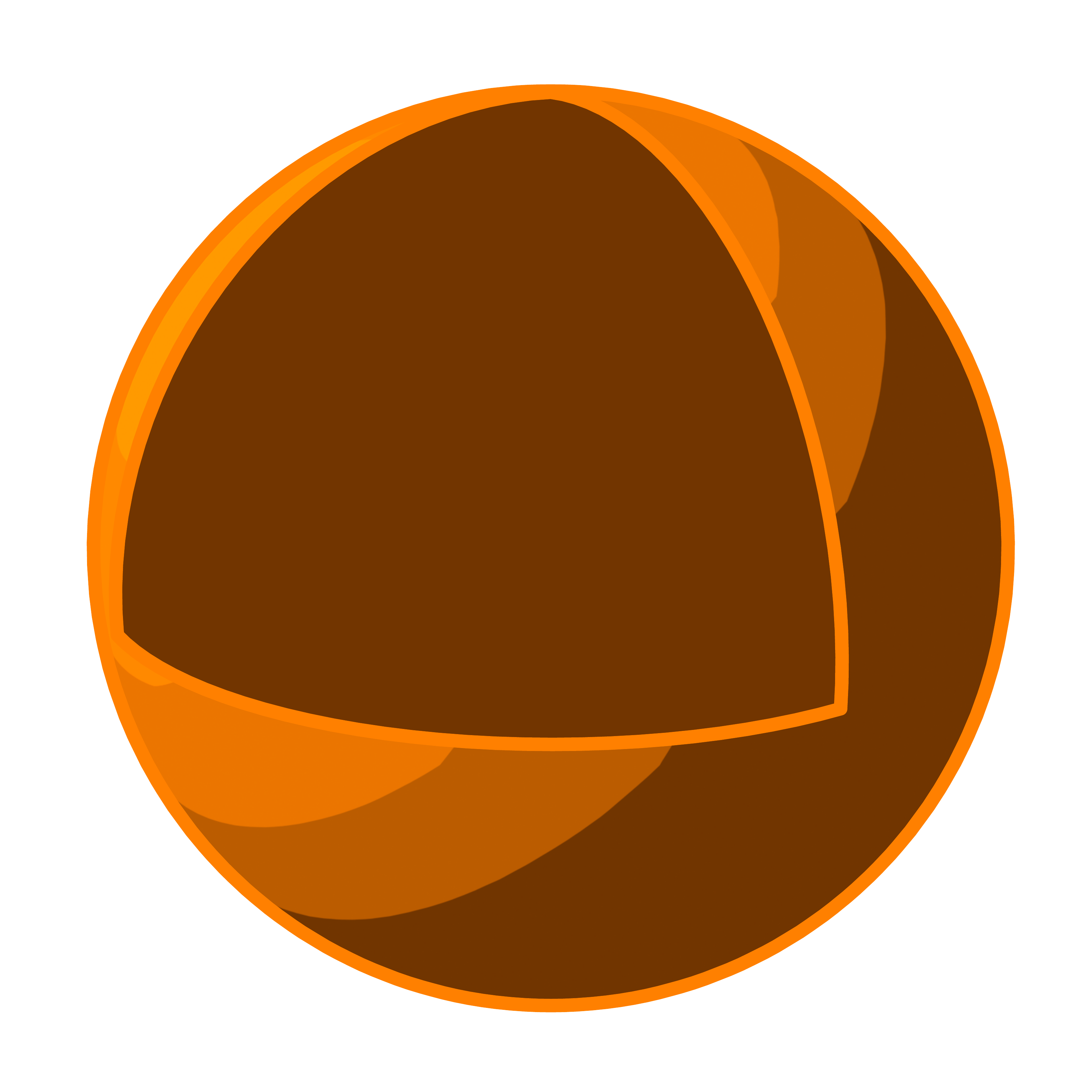}}; 
            \node[below=5mm] at (probe) {Probe};
            
            \node[inner sep=0pt, name=probe1] at (-0.35\linewidth, 0.2) {\includegraphics[width=2cm]{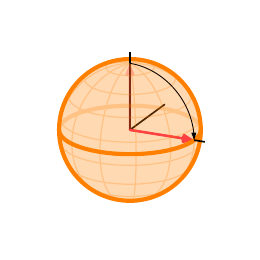}};

            \node[inner sep=0pt, name=probe2] at (0, 0.2) {\includegraphics[width=2cm]{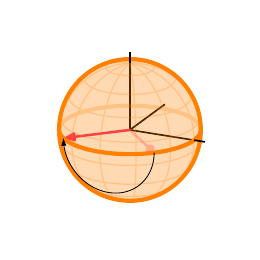}};

            \node[inner sep=0pt, name=probe3] at (0.35\linewidth, 0.2) {\includegraphics[width=2cm]{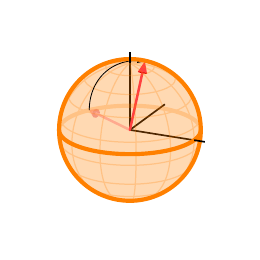}};
        \end{scope}

        \begin{scope}[yshift=-1.1cm] 

            \pgfmathsetmacro{\lw}{\linewidth/1cm}
            \pgfmathsetmacro{\fwhm}{0.22}

            \begin{scope}
                \clip (-0.5\linewidth, -1) rectangle (0.5\linewidth, 1);

                \draw[red!75, fill=red!75, fill opacity=0.5, line width=1pt, domain=-0.5*\lw:0.5*\lw, samples=300] plot[smooth] (\x, {0.4*exp(-(\x+0.35*\lw)*(\x+0.35*\lw)/((0.3*\fwhm)^2)) + 0.4*exp(-\x*\x/(\fwhm^2)) - 1});
            \end{scope}

            \draw[gray, line width=.75pt] (-0.5\linewidth, -1) rectangle (0.5\linewidth, 1); 

            \node[right=2mm] at (-0.35\linewidth, -0.65) {$\alpha$}; 
            \node[right=2mm] at (0, -0.65) {$\pi$}; 

            \node[inner sep=0pt] (source) at (-0.63\linewidth, 0.2) {\includegraphics[width=1cm]{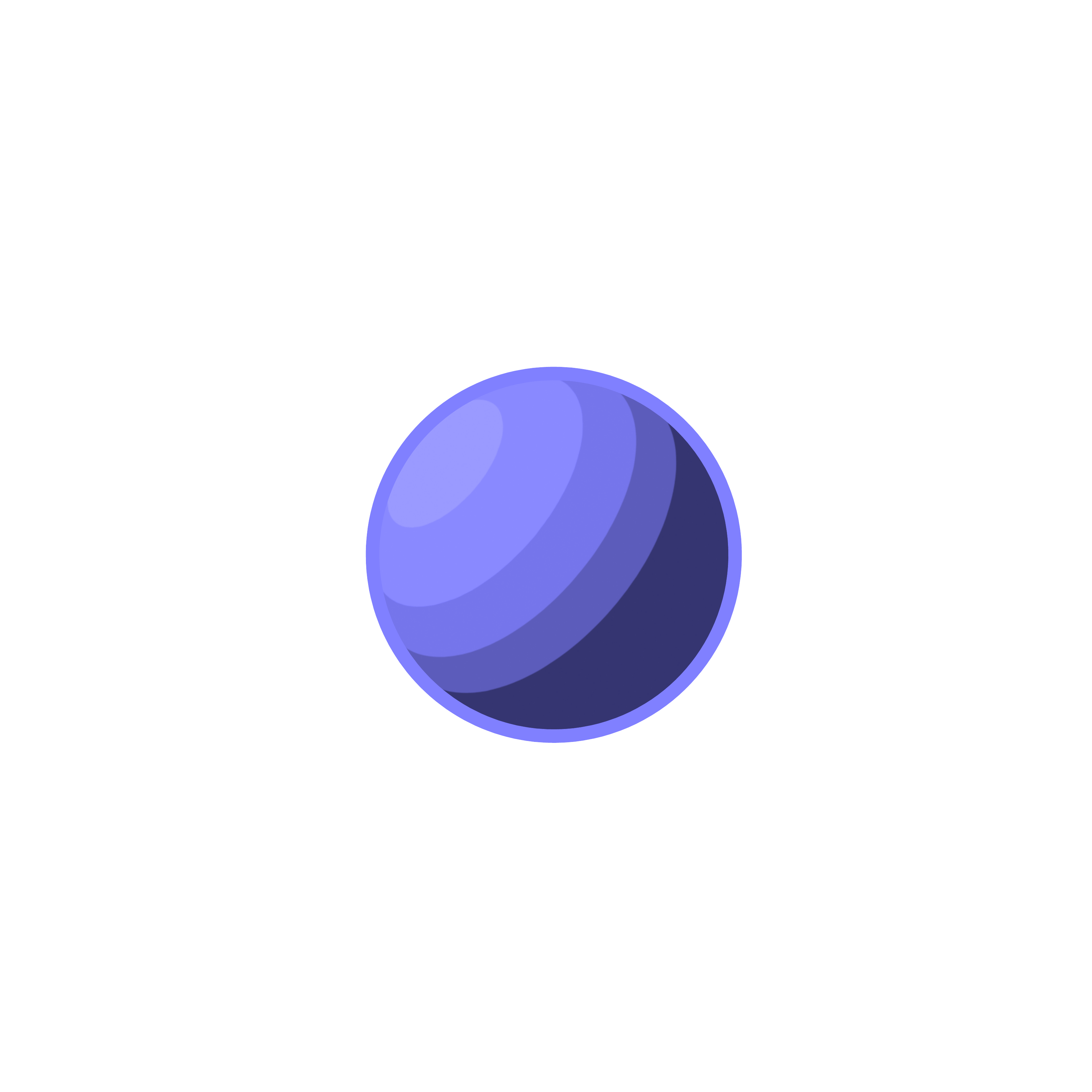}}; 
            \node[below=3mm] at (source) {Source};

            \node[inner sep=0pt, name=source1] at (-0.35\linewidth, 0.2) {\includegraphics[width=2cm]{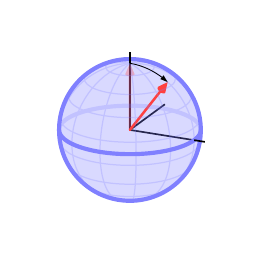}};

            \node[inner sep=0pt, name=source2] at (0, 0.2) {\includegraphics[width=2cm]{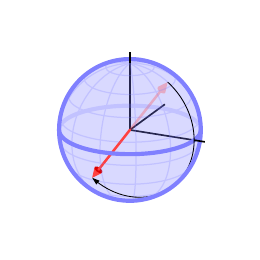}};

            \node[inner sep=0pt, name=source3] at (0.35\linewidth, 0.2) {\includegraphics[width=2cm]{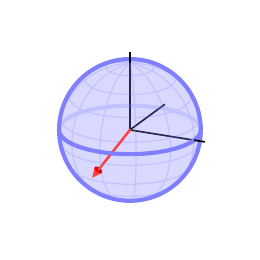}};
        \end{scope}

        \draw[line width=1pt, -latex] (-0.5\linewidth, -2.5) -- (0.5\linewidth, -2.5); 

        \draw[line width=0.5pt] (-0.35\linewidth, -2.4) -- (-0.35\linewidth, -2.6) node[below] {$0$};
        \draw[line width=0.5pt] (0, -2.4) -- (0, -2.6) node[below] {$t_R/2$};
        \draw[line width=0.5pt] (0.35\linewidth, -2.4) -- (0.35\linewidth, -2.6) node[below] {$t_R$};
        
    \end{tikzpicture}
    \caption{Protocol to isolate the gravitational signature sourced by an energy superposition in a nuclear clock. The top row shows the spectroscopic sequence for the probe clock, while the bottom row shows the sequence for the source, represented on a Bloch sphere. The signal depends on the state of the clock through \(\alpha\), which can be isolated from the constant or slowly varying background noise contributions.
    The probe is initialized in the balanced superposition and is flipped by a \(\pi\)-pulse in the middle of the protocol.
    The source is prepared with angle \(\alpha\), i.e.~an energy population proportional to \(p(\alpha) = 1-\cos(\alpha)\) [see Eq.~\eqref{eq:total-energy-mass}]. At time \(t_R/2\) the population is flipped, changing the gravitational redshift seen by the clock. The redshift difference between the first and second part of the protocol is proportional to \(\cos \alpha\). The protocol can be repeatedly applied to generate a time-varying signal and lock-in-like measurement to the specific signal frequency.
    }
	\label{fig:protocol}
\end{figure}

The basic measurement protocol is shown in Fig.~\ref{fig:protocol}. Both the source and the probe clock systems are prepared in a superposition state of ground and excited states of the individual nuclei through application of resonant driving fields for the appropriate duration on both systems. For the probe we choose a balanced superposition, generated by a \(\pi/2\)-pulse. The source nuclei are prepared by a pulse of angle \(\alpha\) generating the state \eqref{eq:initial-product-state}.
After time $t_R/2$, both the source and the probe undergo a \(\pi\)-flip. 
The probe clock superposition state is then recombined at time \(t_R\) by another \(\pi/2\) pulse as in standard Ramsey spectroscopy and is read out. This DEER-like protocol allows for the  cancellation of all constant contributions to the accumulated phase, including the rest mass shift \(\delta_\rest\). 
Due to the population inversion at time \(t_R/2\), the gravitational redshift sourced by the internal energy jumps during the flip, \(\Delta_\avg(\alpha) \mapsto \Delta_\avg(\pi - \alpha)\), and the redshift difference is picked up by the probe nuclei as a phase
\begin{align}
    \phi = \frac{12 G L N_S \hbar \omega_S \omega_P  } {c^{4}(12 L^2 + d^2)} \frac{t_R \cos \alpha}{2},
\end{align}
where \(\omega_S\) and \(\omega_P\) are the frequencies of the source and the probe nuclei, respectively.
Scanning through different initial \(\alpha\) can reveal this sinusoidal signature. 
Sequential periodic application of $\pi$-pulses before the completion of a Ramsey cycle generates a periodic DD-like signal at a controllable frequency, allowing for a rejection of low-frequency noise in the system.

Electromagnetic interactions between the source and the probe are orders of magnitude greater than the gravitational shift and they can themselves depend on \(\alpha\), as is the case for dipole moments that differ in the ground and excited isomer states. 
Electric dipole moments of the nuclear states are strongly suppressed. 
In contrast, the magnetic dipole interaction between the source nuclei and a sensing probe nucleus is significant. 
Placing the probe systems at distance \(r\) from the center of the sourcing sphere with \(N_S\) nuclei, the magnitude of the resulting frequency shift is roughly given by
\begin{align}
    \frac{E_\dipole}{\hbar \omega_P} \sim \mu_0 \frac{N_S \, \tilde\mu_S\,  \tilde{\mu}_P}{r^3 \hbar \omega_P},
\end{align}
where \(\tilde\mu_i = \abs{\mu_i^e - \mu_i^g}\) is the difference in magnetic moment of the excited and ground state of the source and the probe nuclei, respectively.  
The gravitational shift is maximal for the closest possible distance \(r\approx R\). 
Since \(N_S\) scales like \(R^3\) the magnetic dipole shift only depends on the atom number density \(n\) in this case. 
For source and probe made of scandium nuclei with \(\tilde\mu_\text{Sc} = 4.4 \mu_N\) with the nuclear magneton \(\mu_N\)~\cite{shvydkoResonantXrayExcitation2023}, we find
\begin{align}
    \frac{E_\dipole}{\hbar \omega_P} \sim  \frac{4}{3\hbar \omega_P}\pi \mu_0 \tilde\mu_\text{Sc}^2 n= 10^{-13} \, .
\end{align}
Thus this influence is around twenty orders of magnitude larger than the gravitational shift in Eq.~\eqref{eq:Delta-avg-shell}. A larger separation between the source and the probe would in principle improve the ratio between the gravitational and the dipole shift, as these influences fall off differently with distance. However, this would also weaken the gravitational signature itself, make it more prone to other noise sources and increase the fundamentally necessary averaging time.
A successful detection of the gravitational shift will therefore depend on a multitude of mitigation techniques for the magnetic dipole interaction, and amount to ultrahigh-precision measurement of the interaction.

Averaging over the shell of sensing clock atoms can help to cancel multipole electromagnetic effects. 
It has been experimentally demonstrated that for a balanced superposition of the internal state, i.e.\ \(\alpha=\pi/2\), the average dipole contribution vanishes~\cite{hutsonObservationMillihertzlevelCooperative2024}. However, in our case we want to be sensitive to the changing gravitational potential as a function of \(\alpha\), thus necessarily changing also the dipole moment of the inner sphere.
Due to the spherical geometry, the net influence on the clock frequency nevertheless averages to zero, if all dipoles are aligned (see Appendix~\ref{app:coherent-cancelling}).

This average magnetic effect can also be canceled by splitting the probe nuclei ensemble into two hemispheres using nuclear transitions with opposite spin states~\cite{campbellSingleIonNuclearClock2012, beloyTrapInducedAcZeeman2023}. That is, one hemisphere uses the transition \(\ket{0, +m_0} \leftrightarrow \ket{1, +m_1}\), while the other one uses \(\ket{0, -m_0} \leftrightarrow \ket{1, -m_1}\).
In the respective ensembles the gravitational and the magnetic interaction add a phase shift \(
    \phi_\pm(\alpha) = \phi_G(\alpha) \pm \phi_\text{mag}(\alpha)\).
Thus, averaging over the two hemispheres cancels the first-order magnetic shift.
Alternatively, the averaging could be over cycles instead of ensembles~\cite{campbellAbsoluteFrequencyThe872008}, using interleaved alternating measurements on the different states.
Mitigation strategies of this form are used in atomic clocks~\cite{zhengLabbasedTestGravitational2023} and precision measurements of the electron's electric dipole moment \cite{cairncrossPrecisionMeasurementElectrons2017}.
Importantly, the ground and excited isomeric states can have different magnetic moments. Thus, in an applied magnetic field one can use the clock laser frequency and polarization to selectively address different spin state transitions between the two isomer levels in order to isolate magnetic and gravitational interactions.

Given even more advanced control techniques of the nuclei, a  complementary approach can cancel the dipole effects on the probe nuclei altogether. 
One could entangle pairs of nuclei that experience opposite dipole energies (e.g.\ on opposite ends of the sphere) such that the effects compensate each other in the accumulated phase before the read-out (see Appendix~\ref{app:coherent-cancelling} for details). 
Bipartite entanglement is in principle sufficient for such a coherent noise suppression. Alternatively, ensembles with opposite dipole contributions could be spin-squeezed in order to achieve a partial coherent canceling of the effect.

A grounded shield between the two parts could further suppress 
electrical multipole effects.
The shield itself must be well controlled as it can contribute a broadening through vibrations that lead to fluctuations of the induced electromagnetic fields~\cite{bullingStabilityThresholdsGravitationally2026}. However, since we are only interested in 
the frequency shift, the shield does not add genuine quantum noise to the measured signature.
We note that this would drastically change if we tried to generate entanglement through gravity, as envisioned for mechanical systems \cite{boseBMV2017,marlettoVedralBMV2017} (we discuss entanglement between atomic clocks in Appendix~\ref{sec:entanglement}). 
The issue in this case would be the generated entanglement with the shield itself, 
thus, its decoherence would have to be suppressed to the same extent as the measured systems themselves. This makes the use of a macroscopic shield 
problematic in the context of gravitationally induced entanglement, and would require other noise suppression mechanisms that do not contribute to decoherence.

Overall, the outlined techniques show how one might approach the challenge of isolating the gravitational interaction of interest from much larger background noise. It follows in spirit the developed techniques for high-precision measurements in AMO systems, characterizing all possible channels independently. This approach has recently been applied to high-precision measurements in bulk systems \cite{nima2025precision,radak2026sensing}. As nuclear clock technology matures, it may open a path to extract the collective mutual gravitational redshift interaction. An alternative approach to sense the superposition of different energies is to use an external force sensor, as described in the next section.
\medskip

\subsection{Measuring superpositions of the gravitational field with force sensors}\label{sec:force}

\begin{figure}
    \centering
    \begin{tikzpicture}
        \node (a) at (0, 0) {\includegraphics[width=\linewidth]{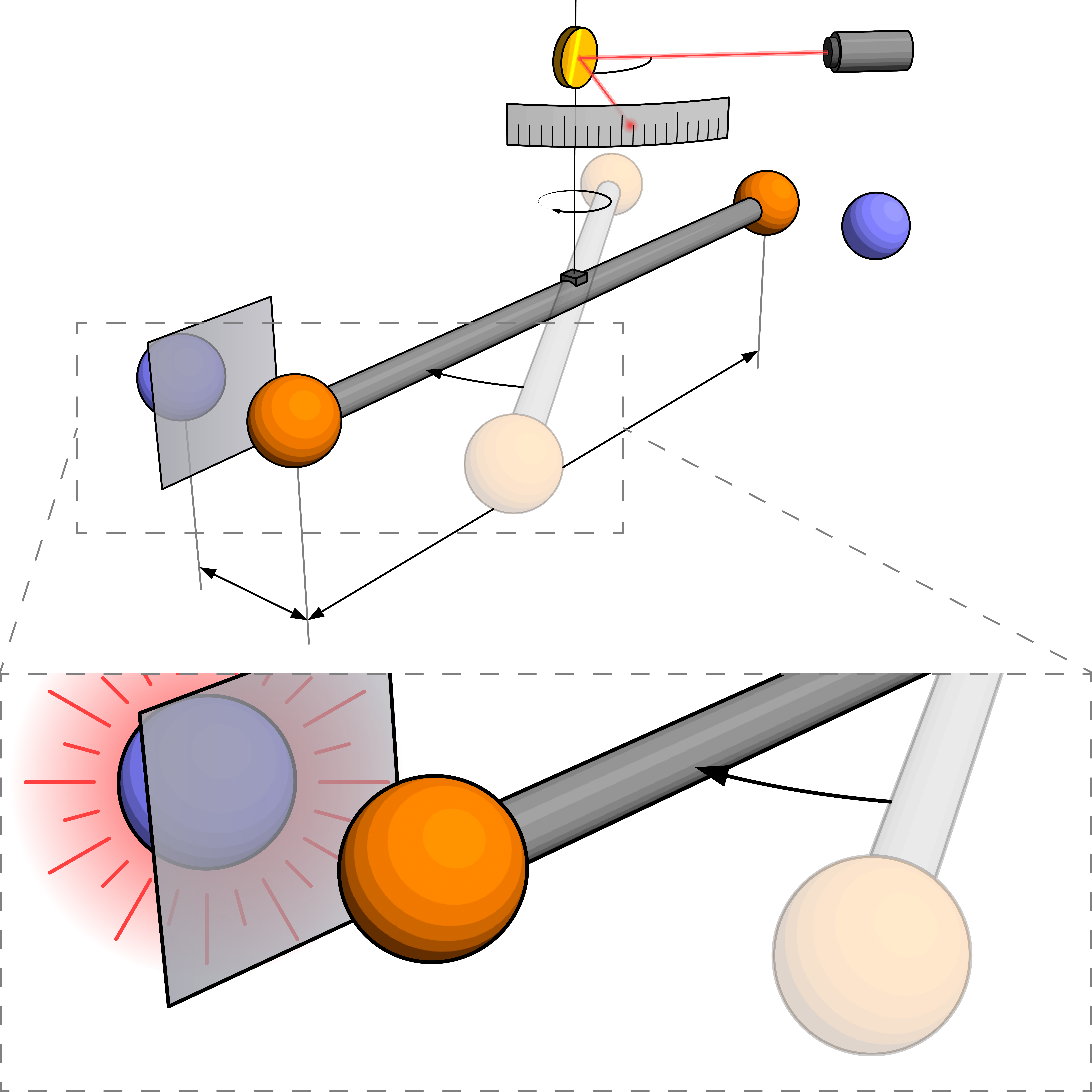}};


        
        
        
        \node[inner sep=2pt] at (-2.55, -3.75) {\footnotesize$\mathcolor{blue!50}{M_0}+ \mathcolor{red!75}{M_E(\alpha)}$};

        \node at (1.5, -2.3) {$\delta\phi(\alpha)$};

        \node[scale=1] at (-0.9, 1) {\footnotesize$\delta\phi$};
        \node at (1.1, 3.7) {\footnotesize$2\delta\phi$};

        \node at (-2.4, -0.6) {\footnotesize$D$};

        \node at (0.2, 0.4) {\footnotesize$L$};
    
    \end{tikzpicture}

        

        





    
    \caption{Torsion balance experiment for the measurement of the active gravitational mass of energy in superposition. The excited state is heavier which leads to an additional deflection of the pendulum. By alternating the excitation and therefore the mass of the clocks through the parameter $\alpha$ [see Eq.~\eqref{eq:initial-product-state}], the deflection angle changes.}
    \label{fig:cavendish}
\end{figure}

Bringing \(N = 6\times 10^{23}\) thorium nuclei from their ground state into the excited state leads to a total change of mass of $10\,\si{\nano\gram}$. 
For scandium, the change would reach $13\,\si{\micro\gram}$, which is of the order of the Planck mass ($m_p \approx 22\,\si{\micro\gram}$). 
This regime is only four orders of magnitude away from the current record for the smallest measured gravitating mass~\cite{westphalMeasurementGravitationalCoupling2021,agafonova2026one}.
Thus, the change of the gravitational mass due to a change of the internal energy might be easier to observe by directly measuring the resulting acceleration of a probe mass,
rather than the gravitational redshift.
Ultra-sensitive sensors to detect the change in the gravitational field could for example be realized with atomic fountains~\cite{asenbaum_phase_2017}, levitated particles~\cite{timberlake2019acceleration,monteiro2020force,xiong2021lens,fuchs_measuring_2024}, microelectromechanical systems (MEMS) \cite{gao2026force}, optomechanical cavities and membranes \cite{guzman2014high, chowdhury2023membrane} or a torsion balance~\cite{schlamminger2008test,westphalMeasurementGravitationalCoupling2021,agafonova2026one,ross2026improved}, as depicted in Fig.~\ref{fig:cavendish}.  
The gravitational acceleration \(a_\ie\) due to the internal energy at a distance \(D\) is given by \(a_\ie = G M_E(\alpha)/D^2\).
To give a quantitative estimate, as the mass changes by $10\,\si{\nano\gram}$, the sensor at distance $D$ would have to resolve a resulting acceleration \(a_E\) of order $10^{-18} \, \mathrm{m \, s^{-2}}$ at centimeter separations.
For torsion balances, a resolution for the measurement of Planck masses is expected in the coming years~\cite{westphalMeasurementGravitationalCoupling2021,agafonova2026one}. 
In order to measure the field of 10\,ng energy-mass in a macroscopic thorium clock, further improvements will be necessary, also because the spatial extension of such a clock sets a lower bound on the possible distance of the force sensor to the center of mass. 

Traditional torsion balance experiments work on the principle of changing the position of a mass in space and measuring the resulting change of deflection of another probe mass.
By changing the internal energy of a solid-state system, here a complementary type of experiment can be realized: instead of relying on multiple settings with different locations of the source mass, the source remains at a fixed location, but its gravitational mass changes as energy is deposited in it.
It is often advantageous to perform such an experiment in a dynamical regime. In our case this would mean that the internal energy is continuously changed by coherently driving the nuclear energy transition up and down, i.e.\ varying $\alpha$. 
Importantly, such measurements are not fundamentally limited by the coherence time of the atoms or the dead time between duty cycles, when using multi-ensemble protocols and interleaving of the clock preparations \cite{zheng2024reducing}. 
The modulation frequency of the clock source can be coherently controlled, and it is advantageous to choose the modulation frequency to coincide with the mechanical resonance of the sensor, \(\omega_\text{mod} = \omega_m\). On resonance the steady-state displacement of a damped harmonic oscillator of mass \(m_m\)
and quality factor \(Q\) is enhanced over its static response,
$x_{\text{peak}} = G M_\ie(\alpha_\text{max}) Q/(D^2 \omega_m^2).$ The sensing strategy is analogous to envisioned measurements of the Newtonian gravitational interaction between milligram-scale masses \cite{schmole2016micromechanical}, but the periodic change in the sourced field would be created by the modulation of the nuclear clock's mass rather than the modulation of its distance to the probe. In both cases, the sensor would detect a periodic gravitational pull, resulting in a modulated acceleration.

In summary, the gravitational field generated by a nuclear clock could be detected with an accelerometer. Comparing the requirement of $10^{-18} \, \mathrm{m \, s^{-2}}$ acceleration sensitivity to current capabilities, LISA Pathfinder demonstrated a differential-acceleration noise floor of approximately $1.7 \times 10^{-15} \, \mathrm{m \, s^{-2}}/\sqrt{\mathrm{Hz}} $ in the few-millihertz band in space \cite{armano2018beyond}. For a coherent sinusoidal acceleration at a known frequency, this corresponds ideally to a sensitivity of approximately $6 \times 10^{-18} \, \mathrm{m \, s^{-2}}$ after one day. Under terrestrial laboratory conditions, compact mechanical accelerometers 
have demonstrated acceleration sensitivities around \(10^{-6} \, \mathrm{m \, s^{-2}/\sqrt{Hz}}\) across frequencies ranging from the hertz to kilohertz regime \cite{timberlake2019acceleration, monteiro2020force, fuchs_measuring_2024, guzman2014high, chowdhury2023membrane}, with some low-frequency MEMS reaching the \(10^{-9} \, \mathrm{m \, s^{-2}/\sqrt{Hz}}\) level \cite{gao2026force}  and levitated submillimeter-sized sensors reaching \(10^{-8} \, \mathrm{m \, s^{-2}/\sqrt{Hz}}\) \cite{xiong2021lens}. 
For differential acceleration measurements, torsion balances operating at millihertz and sub-millihertz modulation frequencies have achieved sensitivities of a few \(10^{-15} \,\mathrm{m \, s^{-2}}\) after long coherent integration \cite{schlamminger2008test}, with recent results reaching below \(10^{-15} \,\mathrm{m \, s^{-2}}\) \cite{ross2026improved}.

Taken together, these results indicate that the superposed gravitational field sourced by a nuclear clock could be sensed by a nearby mechanical accelerometer without requiring sensitivity too far beyond the current state of the art. However, since the development of the nuclear clock is required in any case to create the source, using a second clock of the same type as the probe and applying quantum control for high-precision spectroscopy, as described in the previous section, may ultimately provide a similarly viable alternative.

\medskip

\section{Discussion}
\label{sec:compare}

We have shown above that gravitational interactions between clock atoms may become relevant and measurable in future nuclear clocks with macroscopic atom numbers. 
Importantly, the energy states in the clocks are in general in superposition in these schemes, and thus the effect beyond the rest mass stems from quantum sources of gravity.
More precisely, coherent varying of the energy superposition angle \(\alpha\) in Eq.~\eqref{eq:initial-state-alpha} should result in a sinusoidal behavior of the mass of the system and therefore a corresponding signature [see Eq.~\eqref{eq:delta4-alpha}]. If observed, this would be the first direct measurement of a gravitational field sourced by a superposed mass. We note, however, that the generation of entanglement as a further signature of the quantum interaction is out of reach due to the fundamental decoherence in such systems, as we discuss in more detail in Appendix~\ref{sec:entanglement}.

\subsection{Comparison to other physical platforms}
We now highlight the distinctive features of nuclear clocks as sources of gravity. 
At its core, our scenario involves the storage of energy in nuclear clock transitions, and their contribution to the gravitational field. Expressed in terms of photonic energy, this amounts to storing $N$ photons and probing their resulting gravitational field, as for example proposed in \cite{ratzel2016gravitational}. For a macroscopic number of nuclei, this is a significant amount of energy, which is converted from photons and stored as bound mass-energy in the clocks and then interacts via gravity---which is not feasible for free photons. For energy storage alone, certain metastable nuclear isomers would be advantageous over the thorium, scandium, and uranium transitions discussed above. Most notably, hafnium-178m2 can store about 2.5\,MeV per nucleus with a half-life of 31 years \cite{hartouni2008}. 

However, beyond mere storage, the nuclear clocks allow for the coherent change in the mass, through the superposition parameter $\alpha$.
We can contrast the gravitational fields sourced by superpositions of internal energy with approaches where a superposed gravitational field is sourced by a fixed mass in spatial superposition, such as atomic interferometers \cite{overstreet2022observation,panda2024measuring}. 
To compare these two, a change of the gravitational field sourced by the systems has to be of the same magnitude. As noted above, the superpositions considered in nuclear clocks can be translated into a changing mass of order 10\,ng for Avogadro's number of thorium nuclei. This would amount to a field generated by the rest mass of a cloud of \(10^{14}\) rubidium atoms, each superposed in different locations.
To create product states as in Eq.~\eqref{eq:initial-product-state}, 
a cloud of individually superposed atoms of the same total mass would suffice. 
Nevertheless, observing matter-wave properties of that many atoms remains a formidable challenge.

\subsection{Distinctive properties of energy superpositions}
Sourcing a gravitational field by energy superpositions differs fundamentally from sourcing it by a spatial superposition of rest mass, regardless of whether it is realized in an atom interferometer or alternative platforms such as optomechanical setups.

Since mass-energy equivalence is a relativistic effect, the nuclear clock energy superposition scenario is post-Newtonian. 
Furthermore, mass superpositions in the form of energies generate two branches that differ in their monopole field. This is in contrast to fields sourced by spatially superposed masses where postulated observable effects depend on the difference of the dipole contributions in the branches, since the total mass is the same in both branches.
The possibility of such a monopole superposition is unique to gravity. Superselection rules for the electric charges prevent the creation of superpositions of charge values (superposing the presence and absence of a charge in a certain region is of course allowed but this is again a spatial superposition phenomenon of a single charge).
In contrast, mass and energy do not obey such a superselection rule and can be brought into superposition \cite{Zych2011, zych2019puzzling,giulini_galilei_1996}. A test with nuclear clocks on how energy itself, and in particular mass-energy superpositions, gravitate is thus of particular interest, probing quantum phenomena in the post-Newtonian limit of gravity.

\bigskip

\section{Conclusions}
The rapid advancement of atomic and nuclear clocks opens the possibility of exploring timekeeping in new physical regimes. We have shown that mutual gravitational interactions within clocks---typically negligible at such scales---become notable for the macroscopic atom numbers envisioned for nuclear clocks, with rest mass effects becoming relevant for thorium nuclear systems consisting of \(N \sim 10^{18}\) clock atoms.

Of particular interest is the gravitational field of the mass-energy, a unique post-Newtonian effect that could in principle be detected in future nuclear clocks. While far beyond current experimental capabilities, the rapid development of thorium nuclear clocks together with advances in free-electron laser technology makes such an experiment conceivable. We have shown how the much larger electromagnetic noise floor could in principle be mitigated as AMO high-precision techniques become accessible within nuclear clocks.

Such a setup could reveal the coherent sourcing of gravitational fields from energy superpositions, constituting an important milestone in testing the consistency of quantum mechanics and gravity at low energies---or their breakdown. Our results demonstrate that future nuclear clock technology opens a new platform for probing gravitational interactions on atomic scales, with the potential to reach sensitivities where quantum sources of gravity become relevant.

\section*{Acknowledgements} This work was supported by the National Science Foundation under Grant No. 2239498, by NASA under Grant No. 80NSSC25K7051 and by the “Table-top experiments for fundamental physics” program sponsored by the Alfred P. Sloan Foundation, Gordon and Betty Moore Foundation, John Templeton Foundation, and Simons Foundation under Grant No. G-2023-21102. NEÖ acknowledges funding by the H.~W.~Söderman and F.~O.~Törnlund scholarship foundations as well as the Wallenberg Initiative on Networks and Quantum Information (WINQ). Nordita is funded in part by NordForsk.

\bibliography{bib}

\pagebreak
\appendix

\section{Derivation of the master equation} \label{app:master-equation}

The master equation~\eqref{eq:master-equation} is most easily obtained in a Bloch vector formalism.
The Pauli matrices together with the identity element \(\sigma_0 = \id\) form an orthogonal basis of the real vector space of \(2\times2\) Hermitian matrices.
Thus, defining a vector of matrices \(\boldsymbol\sigma = (\sigma_0,\sigma_x,\sigma_y,\sigma_z)\) we can decompose any qubit density matrix \(\rho\) into the elements
\begin{align}
    k_i = \tr[\rho \boldsymbol{\sigma}_i],\quad i = 0,x,y,z.
\end{align}
We call \(k\) the extended Bloch vector of the state \(\rho\). Note that for a normalized state we have \(k = (1,\Vec{r})\), where \(\vec r\) is the usual three-dimensional Bloch vector in the Bloch sphere.

Any qubit channel \(\mathcal{E}\) can be decomposed into a matrix
\begin{align}
\label{eq:Bloch-map}
    \Lambda_{ij} = \frac{1}{2}\tr[\boldsymbol{\sigma}_i \mathcal{E}(\boldsymbol{\sigma}_j)],
\end{align}
such that the transformation of an extended Bloch vector is given by usual matrix multiplication
\begin{align}
    k' = \Lambda k.
\end{align}

For the dynamics of atom \(A\) under the gravitational influence of atom \(B\) as given in Eq.~\eqref{eq:reduced-dynamics}, we denote the time-dependent Bloch map by \(\Lambda_t\).
For a given input state \(k(0)\) of atom \(A\) it yields the extended Bloch vector at time \(t\):
\begin{align}
    k(t) = \Lambda_t k(0).
\end{align}
In order to derive the master equation, we are interested in the generator of the evolution at any time \(t\).
For the time derivative of the extended Bloch vector we have
\begin{align}
    \dot k(t) = \dot\Lambda_t k(0) = \dot\Lambda_t \Lambda_t^{-1} \Lambda_t k(0) = \dot\Lambda_t \Lambda_t^{-1} k(t).
\end{align}
Thus, the generator at time \(t\) is given by 
\begin{align}
    \xi_t = \dot\Lambda_t \Lambda_t^{-1}.
\end{align}
In order to find the generator in the normal operator form of a pseudo-Lindblad equation, we start from the ansatz in Eq.~\eqref{eq:master-equation}, calculate its generator in the Bloch formalism and compare it to the generator obtained from the dynamics \(\Lambda_t\).
The exact solutions for \(\delta_t\) and \(\Gamma_t\) are given by
\begin{align}
\label{eq:Delta}
    \delta_t \!= \!\frac{G m_0}{c^2 a} \!+ \!\frac{\gamma }{2 \omega_c a}\left( \!1- \frac{2z}{ - (z^2 + 1)  \!+\! (z^2 - 1) \cos(\gamma t/a)} \!\right),
\end{align}
and
\begin{align}
\label{eq:Gamma}
    \Gamma_t = \frac{\gamma(z^2-1)\sin\left(\gamma t/a\right)}{4 a (z^2-1)\cos\left(\gamma t /a  \right) - 4 a (z^2+1)},
\end{align}
with
\begin{align}
\label{eq:gamma}
    \gamma = \frac{G \hbar \omega_c^2}{c^4}\, \quad \text{and} \quad z = \tr[\sigma_z \rho_B],
\end{align}
where \(\rho_B\) is the initial state of atom \(B\).
Both functions are explicitly time-dependent. However, expanding \(\delta_t\) in powers of \(c^{-1}\), the first two relevant terms are constant. For the frequency shift \(\delta_t\), this leads to Eq.~\eqref{eq:single-atom-frequency-shift} in the main text. The first relevant order of the dephasing coefficient \(\Gamma_t\) is given in Eq.~\eqref{eq:depahsing-first-relevant-order}.

\section{Local Dephasing}
\label{app:dephasing}
\subsection{Two atoms}
\label{app:two-atoms-dephasing}
Let us take a look at the second term in Eq.~\eqref{eq:master-equation}. 
It describes a dephasing process in the energy eigenbasis of the local clock Hamiltonian, which by convention is considered the \(z\)-axis here.
The dynamics is non-Markovian, since \(\Gamma_t\) as given in Eq.~\eqref{eq:Gamma} periodically becomes negative which causes the dephasing to decrease again. 
This can be understood by the fact that due to the gravitational interaction Hamiltonian \(H_\ie\) in Eq.~\eqref{eq:Hquantum}, atom \(A\) gets periodically correlated and uncorrelated with atom \(B\) which results in a local dephasing process. 

This dephasing depends on the \(\sigma_z\) expectation value \(z = \tr[\sigma_z \rho_B]\) of the initial state in atom \(B\) but not on its coherences. Thus, on a local level of atom \(A\) alone, the dephasing dynamics caused by an atom \(B\) in equal energy superposition, i.e.\ \(\rho_B = (\id + \sigma_x)/2\) is indistinguishable from the one with atom \(B\) being in a fully mixed state, i.e. \(\rho_B = \id/2\). Both cases correspond to \(z=0\) in Eq.~\eqref{eq:Gamma}.
Thus, the gravitational decoherence due to a mass or energy superposition is the same as the one caused by a classical mixture. 

However, in a standard atomic clock scheme, all atoms are initially prepared in the same pure state with equal energy superposition, i.e., 
\begin{align}
\label{eq:initial-superposition-appendix}
    \ket{\psi_A} = \ket{\psi_B} = \frac{1}{\sqrt{2}} (\ket{0} + \ket{1}).
\end{align}
In this case, the local decoherence can actually be attributed to the entanglement creation between the atoms through the gravitational interaction. 
We define the coherence or visibility of atom \(A\) with respect to the energy basis (which is equivalent to the \(\sigma_z\) basis) by
\begin{align}
    \nu = \sqrt{\tr[\sigma_x \rho_A(t)]^2 + \tr[\sigma_y \rho_A(t)]^2}.
\end{align}
For the initial states given in Eq.~\eqref{eq:initial-superposition-appendix}, we find
\begin{align}
    \nu = \abs{\cos\left(\frac{G \hbar \omega_c^2}{2 c^4 a} t\right)}, \,
\end{align}
consistent with previous work that focused on decoherence due to the mutual redshifts~\cite{castroruizEntanglementQuantumClocks2017}.
We furthermore define the decoherence time as the first zero of \(\nu(t)\), i.e.,
\begin{align}
\label{eq:decoherence-time}
    t_d = \frac{\pi c^4  a}{G \hbar \omega_c^2}.
\end{align}
This is the time it takes for the two atoms to become maximally entangled and therefore locally fully decohered.
The decoherence times for the experimental parameters in Section~\ref{sec:strength} are shown in Table~\ref{tab:decoherence-two-atoms}.

\def\arraystretch{1.5}
\begin{table}[h]
    \centering
    \begin{ruledtabular}
    \begin{tabular}{c|c|c|c}
         & Thorium-229  & Scandium-45  & Uranium-235 \\ \hline
      \(t_d\) [s] \eqref{eq:decoherence-time} &  \(10^{37}\) & \(10^{31}\) & \(10^{35}\)
      \\
       \(T_d \) [s] \eqref{eq:decN} &  \(10^{32}\) & \(10^{26}\) & \(10^{30}\)
    \end{tabular}
    \end{ruledtabular}
    
    \caption{Decoherence times \(t_d\) and \(T_d\) of two and \(N = 6\times 10^{23}\) atoms, respectively, due to their gravitational interaction for system parameters as given in Table~\ref{tab:summary}.
    The time scales for decoherence due to the build-up of entanglement between the atoms is far larger than the age of the universe and therefore the effect is negligible for any realistic experiment.
    }
    \label{tab:decoherence-two-atoms}
\end{table}
These decoherence times exceed the age of the universe by far. Thus, no significant drop in coherence can be expected at reasonable time scales. Furthermore, since decoherence effects are easily masked by various other noise contributions in atomic clocks, this effect is undetectable even with vastly improved experimental capabilities.

\subsection{$N$ atoms}
\label{sec:N-atoms-dephasing}
In the case of \(N\) atoms, the quantum interaction Hamiltonian \(\Hfour\) in Eq.~\eqref{eq:total-Hamiltonian} leads to correlations between any pair of atoms. Thus, a single atom \(i\) locally experiences \(N-1\) independent dephasing processes due to the other atoms. 
We will again assume the standard case of an atomic clock where all atoms are locally brought into an equal energy superposition of their internal degree of freedom, i.e., the initial state of the \(N\) atoms is assumed to be
\begin{align}
\label{eq:initial-product-state-appendix}
    \ket{\psi} = \left[\frac{1}{\sqrt{2}}(\ket{0} + \ket{1})\right]^{\otimes N}. 
\end{align}
This scenario has been studied in Ref.~\cite{castroruizEntanglementQuantumClocks2017}  for the special case where all pairwise distances are approximated by the same value \(r\). This leads to a dephasing effect whose strength scales exponentially in the number of atoms \(N\). 
The case considered here is different and more plausible insofar as it assumes the atoms to be uniformly distributed in space, as is the case in a lattice clock. Since the strength of the gravitational interaction drops with distance, the total effect scales differently in \(N\) than in the case considered in Ref.~\cite{castroruizEntanglementQuantumClocks2017}.

In Appendix~\ref{app:two-atoms-dephasing}, we have seen that in the two-atom case the visibility of each individual atom goes like
\begin{align}
    \nu = \abs{\cos\left( \frac{\gamma}{2a}t\right)},
\end{align}
where \(\gamma\) as in Eq.~\eqref{eq:gamma} and \(a\) denotes the distance between the two atoms.
Since the dephasing channels in the same basis commute, the visibility of a single atom \(i\) in the grid is then given by the product of the individual contributions, i.e.,
\begin{align}
\label{eq:visibility-N}
    \nu_i = \prod_{j\neq i} \abs{\cos\left( \frac{\gamma}{2r_{ij}}t\right)}.
\end{align}
A closed form of this product cannot be given for the mutual distances \(r_{ij}\) between the atoms on the grid. 
We can nevertheless calculate an estimate of the strength of the dephasing effect, and in particular its scaling with \(N\).

Let us assume that the atoms interact for a duration \(t\).
In order to be observable, the visibility \(\nu\) must drop significantly in this time interval.
We expand Eq.~\eqref{eq:visibility-N} in \(t\). The first order is zero. To second order we find
\begin{align}
\label{eq:V-i-N}
    \nu_i = 1 - t^2 \frac{\gamma^2}{8} \sum_{j\neq i} \frac{1}{r_{ij}^2} + O\left(t^4\right).
\end{align}
Similar to the geometric factor for the frequency shift in Eq.~\eqref{eq:Xi-i}, we define the dimensionless quantity
\begin{align}
\label{eq:Upsilon-i}
    \Upsilon_i = a^2 \sum_{j\neq i} \frac{1}{r^2_{ij}}.
\end{align}
The dephasing effect becomes relevant if the second term in Eq.~\eqref{eq:V-i-N} is of order \(1\),
which yields the following relation between the total time \(t\) and the other experimental parameters:
\begin{align}
     \frac{\gamma^2 t^2}{8a^2} \Upsilon_i = 1.
\end{align}
We define 
\begin{align}
\label{eq:decoherence-time-N}
    T_d^i = \frac{2a}{\gamma}\sqrt{\frac{2}{\Upsilon_i}},
\end{align}
which is roughly the decoherence time for given \(\gamma\), \(a\), and \(\Upsilon_i\) for atom \(i\).

Let us assume a spherical symmetry of the atomic ensemble and consider an atom \(i\) at the center since this will be the one that experiences the strongest decoherence. 
We approximate the sum in Eq.~\eqref{eq:Upsilon-i} by an integral over the atom density \(\atomdensity\) and find
\begin{align}
    \Upsilon_\text{center} = 4 \pi \atomdensity a^2 R,  
\end{align}
where \(R\) is the radius of the sphere.
Using Eqns.~\eqref{eq:radius} and \eqref{eq:decoherence-time-N}, we have
\begin{align} \label{eq:decN}
    T_d^\text{center} =   \left(\frac{32}{3\pi^2 \atomdensity^2} \right)^{1/6} \frac{c^4}{G \hbar \omega_c^2} N^{-1/6}
\end{align}
In Table~\ref{tab:decoherence-two-atoms}, we show this decoherence time for \(N = 6\times 10^{23}\). Even for this macroscopic number of atoms, entanglement generation and the resulting local decoherence occur on timescales far longer than the age of the universe.

\section{Entanglement-assisted noise cancellation}
\label{app:coherent-cancelling}

\begin{figure}
    \centering
    \begin{minipage}{0.49\columnwidth}
        \centering
        \begin{tikzpicture}
            \node[inner sep=0pt] at (0, 0) {\includegraphics[width=\columnwidth]{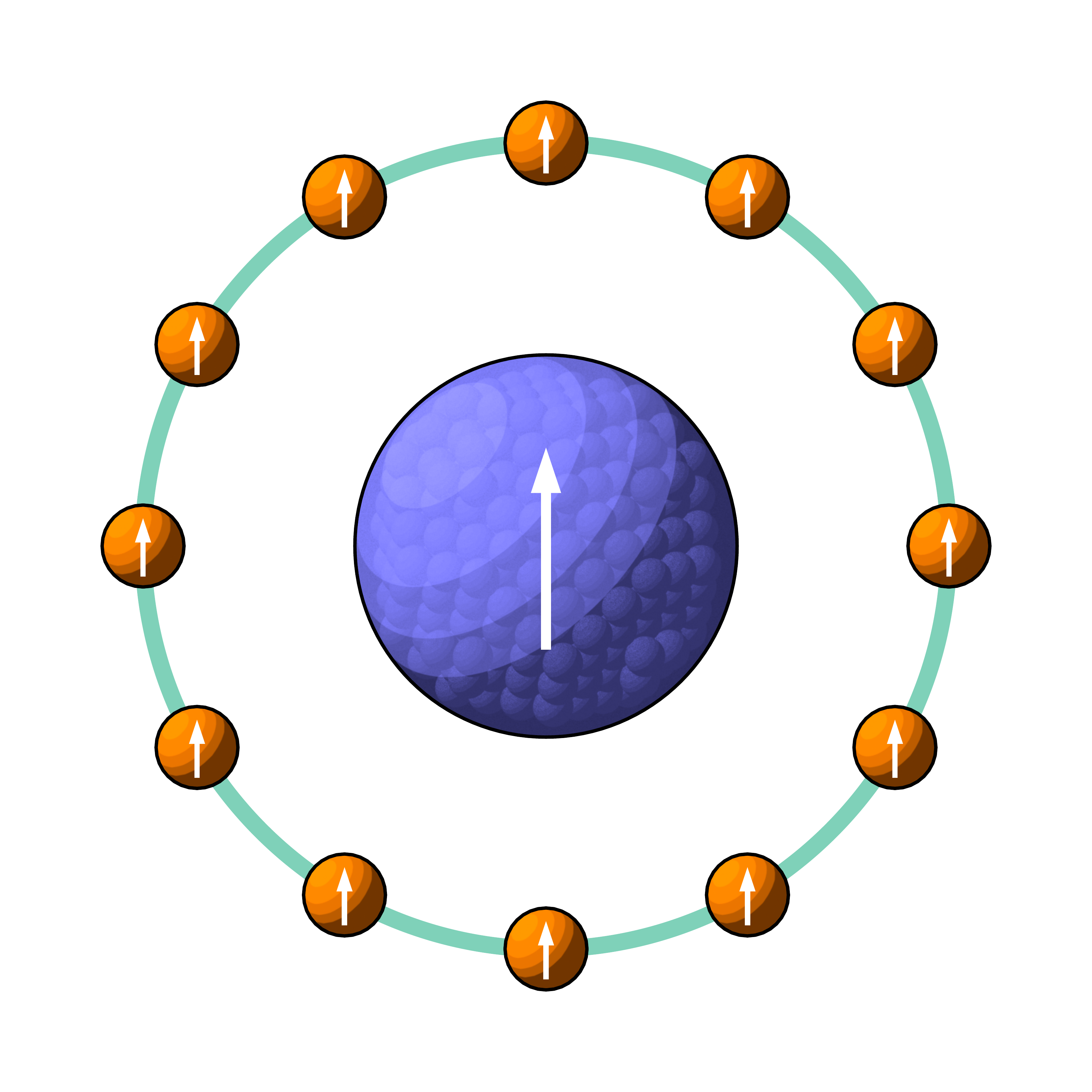}};
            \node[below right] at (-0.5\columnwidth, 0.5\columnwidth) {(a)};
            \node at (-1.5, 1.15) {$\vec p$};
            \node[white] at (0.25, 0) {$\vec P$};
        \end{tikzpicture}
    \end{minipage}\hfill
    \begin{minipage}{0.49\columnwidth}
        \centering
        \begin{tikzpicture}
            \node[inner sep=0pt] at (0, 0) {\includegraphics[width=\columnwidth]{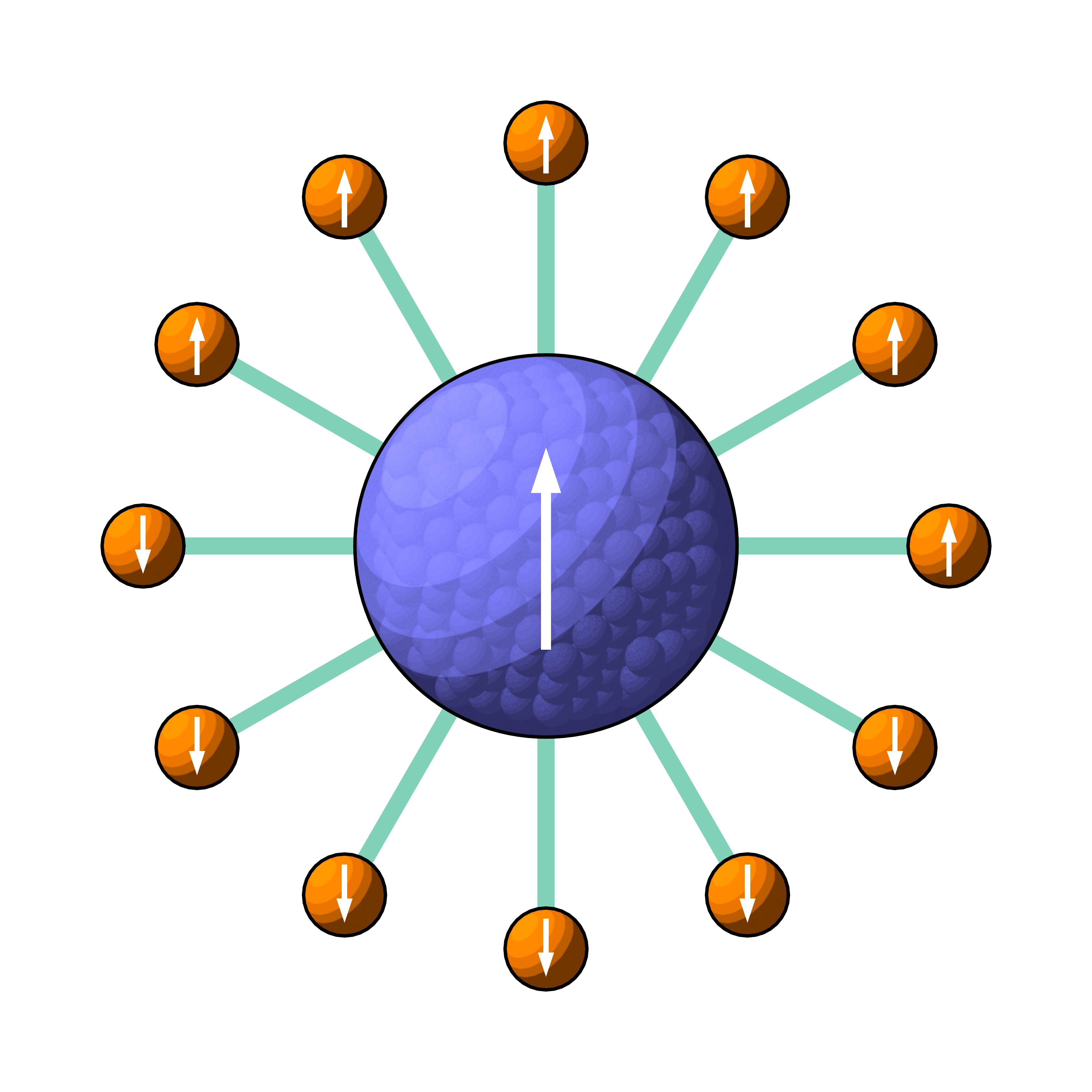}};
            \node[below right] at (-0.5\columnwidth, 0.5\columnwidth) {(b)};
            \node at (-1.5, 1.15) {$\vec p_1$};
            \node at (1.5, -1.15) {$\vec p_2$};
            \node[white] at (0.25, 0) {$\vec P$};
        \end{tikzpicture}
    \end{minipage}
    \caption{We want to detect the monopole contribution of the gravitational field of the internal energy by measuring the redshift of the clock nuclei in the shell (orange) which is sourced by the sphere in the center (blue). Electromagnetic dipole contributions cancel on average if the shell is spherical and has a uniform dipole alignment (a). However, the variance of the dipole interaction energies does not vanish, which leads to an inhomogeneous broadening that may mask the gravitational signal. 
    Using a GHZ state, the broadening can in principle be canceled coherently (see main text). 
    A more practical alternative scenario is to use only bipartite entangled states in order to cancel dipole contributions coherently.
    Aligning the dipoles of the shell nuclei on the two hemispheres in opposite directions (b) and entangling pairs of nuclei on antipodal positions (green lines), the broadening cancels coherently. 
    }
    \label{fig:coherent-cancelling}
\end{figure}

Here we analyze the influence of electromagnetic dipole-dipole interactions on the signal in the shell configuration depicted in Fig.~\ref{fig:shell-setup}. 
Let us assume that all dipoles of individual nuclei/atoms are aligned and point in the \(z\) direction. 
For symmetry reasons, the resulting total dipole of the sourcing sphere in the center can be replaced by a single dipole \(\vec P\) of equal magnitude at the center of the sphere (see Fig.~\ref{fig:coherent-cancelling}).
The energy of an individual dipole \(\vec p\) in the sensing shell is then shifted by a dipole energy of the form
\begin{align}
    E_\dipole = \kappa \frac{\vec p \cdot \vec P - 3(\vec p \cdot \vec n) (\vec P \cdot \vec n) }{r^3}
\end{align}
where \(\vec p\) and \(\vec P\) are the dipoles (electric or magnetic) of a clock nucleus in the outer shell and of the sourcing sphere, respectively. The displacement vector is decomposed into magnitude and direction as \(\vec r = r \vec n \), and the pre-factor \(\kappa\) depends on the nature of the dipole (\(\kappa = 1/4\pi \varepsilon_0\) in the electric case, \(\kappa = \mu_0/4\pi\) for magnetic interactions).
In our case of aligned dipoles and a constant distance \(L\) of the shell nuclei from the center (for ease of calculation we assume a negligible thickness of the shell here, but the same reasoning works for finite shells too), the expression simplifies to
\begin{align}
    E_\dipole = \kappa \frac{p P}{L^3}(1-3 \cos^2 \theta).
\end{align}
The mean value over the whole sensing shell vanishes,
\begin{align}
\label{eq:average-dipole}
   \overline{E}_\dipole &= \frac{1}{A} \int \!\! \diff A\, E_\dipole \notag \\
   &= \frac{1}{4 \pi L^2} \int_0^{2\pi}\!\! \diff \phi \int_0^{\pi} \!\! \diff \theta \, E_\dipole L^2 \sin \theta  = 0.
\end{align}
Thus, for the given geometry, dipole interactions do not contribute to a frequency shift of the sensing clock.
However, due to the different dipole energies of individual clock nuclei, the interaction leads to an inhomogeneous broadening of the signal. 
The variance of the energy (and therefore phase contribution) is given by
\begin{align}
    \operatorname{var}&(E_\dipole) = \frac{1}{A} \int \!\diff A\, E^2_\dipole \notag\\
    &= \frac{1}{4\pi L^2}\left(\frac{\kappa p P}{L^3}\right)^{\!2} \int_0^{2\pi}\!\! \diff \phi \int_0^{\pi} \!\! \diff \theta (1-3 \cos^2\theta)^2 L^2 \sin \theta \notag\\
    &= \frac{4 \kappa^2 p^2 P^2}{5 L^6}.
\end{align}
Since the electromagnetic dipoles are much stronger in magnitude than any gravitational contribution, this broadening can easily mask the gravitational shift of the average frequency we are looking for.
In principle, this broadening can be canceled entirely with the help of entangled states as we describe now.

\subsection{Global coherent cancellation}
Since the dipole energy vanishes on average, a conceptually straightforward way to cancel its contribution makes use of a global GHZ state of all sensing nuclei
\begin{align}
\label{eq:GHZ-coherent-appendix}
    \ket{\psi} = \frac{1}{\sqrt{2}} \left( \ket{0}^{\otimes N} + \ket{1}^{\otimes N} \right).
\end{align}
Such a state would pick up the phase
\begin{align}
\ket{\psi(t)} = \frac{1}{\sqrt{2}} \left( \ket{0}^{\otimes N} + e^{-i (\phi_c + \phi_g + \phi_\dipole)} \ket{1}^{\otimes N} \right),
\end{align}
where \(\phi_c = N \omega_c t\) is the \(N\)-fold phase shift due to the free clock evolution, \(\phi_g\) is the total gravitational shift 
\begin{align}
    \phi_g = -N \omega_c \Delta_\avg  t,
\end{align}
(under the assumption that all \(N\) nuclei see approximately the same redshift \(\Delta_\avg\) as given in Eq.~\eqref{eq:Delta-avg-shell}), and \(\phi_\dipole\) is the total dipole contribution given by the sum of all individual contributions
\begin{align}
\phi_\dipole = \frac{1}{\hbar}\sum_k E_\dipole^k t = 0.
\end{align}
As we have shown above, this contribution vanishes for the spherical configuration (and assuming \(N \gg 1\) in order for the integral to be an appropriate approximation).
Thus, for a GHZ state dipole effects do not contribute to the total phase shift.
While conceptually elegant, the use of a globally entangled GHZ state is unrealistic in an experimental setup (see also Appendix~\ref{sec:entanglement}). 
However, by changing the dipole alignment in the shell, it is possible to coherently cancel the effect with bipartite instead of global entanglement, as we show next.

\subsection{Bipartite coherent cancellation}
The vanishing average dipole interaction energy in Eq.~\eqref{eq:average-dipole} is a global effect encoded in the full integral over a spherical configuration. 
In order to cancel the dipole contribution with bipartite entanglement, we need a scenario where pairs of nuclei see the same dipole energy with opposite sign. 
We achieve this by aligning the dipoles of the northern and southern hemisphere in opposite directions (see Fig.~\ref{fig:coherent-cancelling}\,b).
Clock nuclei at opposite locations in the shell see opposite energy shifts and therefore opposite phase shifts. 
Starting in an entangled state of such a pair of nuclei $A$ and $B$,
\begin{align}
    \ket\psi = \frac{1}{\sqrt{2}}\left(\ket{00} + \ket{11}\right),
\end{align}
we again get a state of the form
\begin{align}
    \ket\psi = \frac{1}{\sqrt{2}}\left(\ket{00} + e^{-i (\phi_c + \phi_g + \phi_\dipole)}\ket{11}\right).
\end{align}
For the case of two entangled nuclei, the free clock evolution leads to \(\phi_c = 2\omega_c t\). 
As before the gravitational monopole contribution is the same for both nuclei and we have \(\phi_g = -2 \omega_c\Delta_\avg t\).
However, the two dipole contributions cancel coherently due to their opposite sign:
\begin{align}
\phi_\dipole = \frac{t}{\hbar}(E_{\dipole}^A + E_{\dipole}^B) = 0.
\end{align}
Thus, by tweaking the alignment of the involved dipoles we can cancel their contribution coherently with only bipartite entanglement. 
While the preparation of such states is  challenging, it is much more practical than the globally entangled GHZ state.
More importantly, this strategy does not rely on a global cancellation, which is only valid in the continuous limit and the very specific spherical geometry. 
The bipartite strategy works in much simpler settings as long as a pairwise matching of opposite contributions can be engineered. 
Therefore, it may find application in many other settings where monopole signals must be isolated from a dipole background.

\section{Practical limitations on entanglement generation through gravity}
\label{sec:entanglement}

A measurement of the gravitational attraction of a superposed mass in nuclear clocks would certainly demonstrate the expected physics of how superposed energy gravitates.
However, even though the frequency shift \(\Delta_\ie\) due to the internal energies theoretically emerges from the quantum interaction Hamiltonian \(H_\ie\) in Eq.~\eqref{eq:Hquantum}, the shift itself is not an unambiguous quantum gravitational signature. 
This can be seen from the fact that the strength of the shift in atom \(A\) depends only on the average energy \(\langle E_B \rangle\) in atom \(B\) [see Eq.~\eqref{eq:delta-qu}].
Thus, the frequency shifts are readily explainable by a mean field model for gravity and therefore do not unambiguously distinguish quantum features of gravity from a semi-classical description.

In Section~\ref{sec:time-evolution} we have briefly mentioned that the quantum interaction Hamiltonian \(H_\ie\) can in principle create entanglement between individual clock atoms and that this entanglement would show up as a decoherence process in the subsystems, as proposed in Ref.~\cite{castroruizEntanglementQuantumClocks2017}. 
However, as shown in Appendix~\ref{app:dephasing}, this effect is completely negligible 
even for macroscopic atom numbers \(N\). 

In principle, we could obtain much stronger entanglement signatures if we used collective effects of correlated initial states. However, the fundamentally increased decoherence rate makes such an experiment practically impossible as we will show now for the case of entanglement between two nuclear clocks as well as between a clock and a mass in spatial superposition.

\subsection{Two clocks in GHZ states}
Preparing the atoms in each clock in a correlated energy superposition given by a Greenberger-Horne-Zeilinger (GHZ) state of ground and excited state
\begin{align}
\label{eq:GHZ}
    \ket{\psi} = \frac{1}{\sqrt{2}} \left( \ket{0}^{\otimes N} + \ket{1}^{\otimes N} \right),
\end{align}
the mass difference between the two branches is
\begin{align}
\label{eq:mass-diff}
    M_\ie = \frac{N \hbar \omega_c}{c^2}.
\end{align}
For instance, a thorium clock with \(N = 10^{23}\) nuclei in a GHZ state would constitute a macroscopic superposition of mass in the nanogram range.
Two such clocks 
would then entangle rapidly due to the superposition of the gravitational potentials associated with the mass superpositions.

However, although complementary to optomechanical implementations of gravitationally induced entanglement (GIE) based on spatial superpositions of masses~\cite{boseBMV2017,marlettoVedralBMV2017}, this clock-based GIE setup is much less feasible.
The preparation of such a macroscopic GHZ state would likely be much harder than the mass superpositions in different positions---although these also require significant improvement over the current record of \(10^{-22}\) \,kg~\cite{pedalino_probing_2026}. While the preparation of the collective quantum states of the clocks could in principle be envisioned through collective spin squeezing, the real limitation will be the decoherence from spontaneous emission, which will be more detrimental for energy superpositions than for spatial superpositions. 
The reason is that in a clock of \(N\) atoms we can roughly assume that a GHZ state decoheres with a rate that scales like \(\gamma_N = N \gamma = N/\tau\), where \(\tau\) is the lifetime of the energy transition. 
Even at absolute zero temperature this lifetime \(\tau\) is finite. Suppressing spontaneous decay would require engineering of a photonic band gap or similar techniques. While such techniques to suppress spontaneous emission have been proposed and implemented in specific systems~\cite{kleppner_inhibited_1981, sandner_spatial_2011, sanner_pauli_2021, jakubczyk_inhibition_2014}, they would have to be applied to the macroscopic clock system, which is at least currently infeasible.

Let us provide a quantitative estimate of the entanglement rate in such a setup with two clocks in large GHZ states. It is effectively given by
\begin{align}
    \omega_\ent = \frac{G M_\ie^2}{\hbar r} = \left(\frac{4\pi \atomdensity}{3}\right)^{1/3}\frac{G \hbar \omega_c^2 N^{5/3}}{ c^4},
\end{align}
where we have assumed the concentric setup in Fig.~\ref{fig:shell-setup} with no distance between the inner sphere and the shell. Due to the scaling with \(N^{5/3}\), there is in principle a regime where the entanglement can beat the decoherence \(\gamma_N\) for the GHZ case. 
Solving for \(N\) we find
\begin{align}
    N > \sqrt{\frac{3}{4\pi}} \frac{c^6}{\omega_c^3} \left(\frac{a}{G \hbar \tau} \right)^{3/2}.
\end{align}
Using the values in Table~\ref{tab:summary} for 
thorium, scandium, and uranium we get, respectively:
\begin{align}
    N^\text{(Th)} \approx 10^{50}, && N^\text{(Sc)}\approx 10^{45}, && N^\text{(U)} \approx 10^{46}.
 \end{align}
Thus, even though entanglement generation could in principle exceed the decoherence, the required atom numbers are beyond physically relevant scales.
Hence, the measurement of gravitational entanglement between nuclear clocks  is out of reach, unless nuclear species can be used with exceptionally small decay rates and large energy transitions (such as in the 2.5\,MeV hafnium-178m2 isomer with a half-life of 31 years~\cite{hartouni2008}), or if the spontaneous decay is significantly reduced by engineering the density of states of vacuum.

\subsection{A clock entangling with a mechanical degree of freedom}
Replacing one of the clocks by a spherical mass in spatial superposition can in principle increase the entanglement rate due to the higher mass density compared to the internal energy mass. 
Assuming the idealized case where one branch of the spatial superposition is as close as possible to the clock and the other branch is at infinity, we find
\begin{align}
\label{eq:ent-rate-mechanical}
\omega_\ent = \frac{G N \hbar \omega_c  m_B}{\hbar c^2 (R_\cl + R_B)},
\end{align}
where \(m_B\) is the mass of the sphere in superposition, \(R_B\) is its radius and \(R_\cl\) is the radius of the nuclear clock which depends on the atom number \(N\) (see Eq.~\eqref{eq:radius}). 
For simplicity, we assume that the decoherence of the mass \(m_B\) in superposition is negligible and only consider the fundamentally limiting spontaneous emission 
of the internal clock states. 
In order for the entanglement rate to exceed this decoherence we again need \(\omega_\ent > \gamma_N = N/\tau\), which together with Eq.~\eqref{eq:ent-rate-mechanical} leads to the condition
\begin{align}
    \frac{G \omega_c m_B}{c^2\left[a\left(\frac{3}{4\pi}N \right)^{1/3} + R_B\right]} > \frac{1}{\tau},
\end{align}
where we have inserted the radius of the clock given in Eq.~\eqref{eq:radius}.
Unlike in the case of two clocks, increasing the number of clock atoms \(N\) does not help and entangling a nuclear clock with a mass in spatial superposition is not a viable way to demonstrate gravitationally induced entanglement. To give an example, using the thorium parameters given in Table~\ref{tab:summary}, \(N=1\), and a clock radius \(R_\cl \approx 0\), the mass of a superposed gold sphere would need to satisfy $m_B \sim 10^{10}~$kg to generate entanglement, which is not a realistic target for a mass in superposition.

\subsection{Decoherence mitigation}
The analysis above shows that spontaneous emission of internal energy levels imposes a significant limitation on gravity-mediated entanglement in both the inter-clock and clock–force-sensor configurations, placing either protocol far beyond present experimental reach. However, this is not a fundamental no-go result. The lifetime depends on the internal transition used and also on the electromagnetic density of states. The latter could in principle be reduced through vacuum-mode engineering, for example using cavities or photonic-band-gap structures designed to inhibit emission. A complementary possibility is to employ clock states with intrinsically much longer lifetimes. In bosonic $^{88}\mathrm{Sr}$, the ${}^{1}S_{0}\!-\!{}^{3}P_{0}$ transition lacks the hyperfine-induced coupling present in fermionic \(^{87}\mathrm{Sr}\), giving the unperturbed excited state an intrinsic lifetime of potentially \(10^{3}\) years \cite{santra2004properties}, compared with the measured \(^{87}\mathrm{Sr}\) lifetime of approximately \(167\) s \cite{dolde2025direct}. The otherwise forbidden transition can nevertheless be driven through a controllable, weak magnetic-field-induced admixture \cite{taichenachev2006magnetic}. Such strategies could substantially relax the decoherence constraint, although achieving the required suppression without introducing other limiting channels would demand major future experimental advances.

\end{document}